\pdfoutput=1 
\documentclass[fleqn,usenatbib]{mnras}

\usepackage{newtxtext,newtxmath}

\usepackage[T1]{fontenc}
\usepackage{ae,aecompl}
\usepackage{hyperref}
\usepackage{comment}

\usepackage{graphicx}	
\usepackage{amsmath}	

\usepackage{amssymb}	
\usepackage{subcaption}
\usepackage{csvsimple}
\usepackage{pgfplotstable,booktabs}
\usepackage{makecell}

\title[LOFAR GW follow-up]{Searching for low radio-frequency gravitational wave counterparts in wide-field LOFAR data}

\author[K. Gourdji et al.]{
K. Gourdji,$^{1}$\thanks{E-mail: \href{mailto:k.gourdji@uva.nl}{k.gourdji@uva.nl} (KG)}
A. Rowlinson,$^{1,2}$
R.~A.~M.~J. Wijers,$^{1}$
J. W. Broderick,$^{3,2}$\newauthor
\,\,A. Shulevski$^4$ and
P.~G. Jonker$^{5,6}$
\\
$^{1}$Anton Pannekoek Institute for Astronomy, University of Amsterdam, Science Park 904, 1098 XH Amsterdam, The Netherlands\\
$^{2}$ASTRON, Netherlands Institute for Radio Astronomy, Oude Hoogeveensedijk 4, 7991 PD Dwingeloo, The Netherlands\\
$^{3}$International   Centre   for   Radio   Astronomy   Research,Curtin   University,   GPO   Box   U1987,   Perth, WA 6845, Australia\\
$^{4}$Leiden Observatory, Leiden University, PO Box 9513, NL-2300 RA Leiden, The Netherlands\\
$^{5}$Department of Astrophysics/IMAPP, Radboud University Nijmegen, P.O. Box 9010, 6500 GL, Nijmegen, The Netherlands\\
$^{6}$SRON, Netherlands Institute for Space Research, Sorbonnelaan 2, 3584 CA, Utrecht, The Netherlands}
\date{Accepted XXX. Received YYY; in original form ZZZ}

\pubyear{2021}

\begin{document}
\label{firstpage}
\pagerange{\pageref{firstpage}--\pageref{lastpage}}
\maketitle

\begin{abstract}
The electromagnetic counterparts to gravitational wave (GW) merger events are highly sought after, but difficult to find owing to large localization regions. In this study, we present a strategy to search for compact object merger radio counterparts in wide-field data collected by the Low-Frequency Array (LOFAR). In particular, we use multi-epoch LOFAR observations centred at 144\,MHz spanning roughly 300\,deg$^2$ at optimum sensitivity of a since retracted neutron star--black hole merger candidate detected during O2, the second Advanced Ligo--Virgo GW observing run. The minimum sensitivity of the entire (overlapping) 1809\,deg$^2$ field searched is 50\,mJy and the false negative rate is 0.1 per cent above 200\,mJy. We do not find any transients and thus place an upper limit at 95 per cent confidence of 0.02 transients per square degree above 20\,mJy on one, two and three month timescales, which are the most sensitive limits available to date. Finally, we discuss the prospects of observing GW events with LOFAR in the upcoming GW observing run and show that a single multi-beam LOFAR observation can probe the full projected median localization area of binary neutron star mergers down to a median sensitivity of at least 8\,mJy.

\end{abstract}




\section{Introduction}
The landmark detection of gravitational waves (GWs) from binary neutron star (BNS) merger GW170817 \citep{abbott17GW} and of its electromagnetic (EM) counterpart spanning the full spectrum has taken us fully into the era of multi-messenger astronomy \citep{abbott17EM}. The rich observational dataset of this single event proved to be a trove of scientific breakthroughs affecting many aspects of astronomy \citep[e.g.][]{abbott17H0,MargalitMetzger17,Cowperthwaite17}. Given the unparalleled impact from the extensive multi-wavelength follow-up campaign of GW170817, there is wide interest in pursuing the EM counterparts of other GW merger events involving at least one neutron star, and particularly of a black hole and neutron star (BH-NS) for the first time \citep[e.g.][]{Dobie19,Coughlin2019,Antier2020,Andreoni2020,Page2020,Alexander2021,Boersma2021}. However, this endeavour comes with substantial challenges, due in large part to the significant uncertainties on the location of GW events: the median 90 per cent credible region of the sky localization area in the previous Advanced Ligo--Virgo GW observing run (O3) was approximately 200\,deg$^2$ \citep{abbott2020prospects}. While event localisation improves with each subsequent GW observing run, large survey speeds and telescopes with large fields of view in particular will remain instrumental in efforts to identify GW EM counterparts in the upcoming observing run (O4), which is likely to commence in the latter half of 2022.

Radio interferometers are particularly promising tools for identifying the exact location of a GW event, owing to their combined large fields of view and sensitivity, as well as low-latency observing capability. The interaction of the relativistic jet that is launched in the moments following a BNS (and possibly BH-NS) merger with the circum-merger medium produces a radio afterglow that is detectable days to weeks post-merger at GHz frequencies and lasting months to years \citep[][]{PaczynskiRhoads93,MeszarosRees97}. This emission was studied at unprecedented detail with radio observations of GW170817 \citep{abbott17EM,Hallinan17}. This synchrotron emission contains key information about the enigmatic jet that produced it and can be traced across the radio band as it evolves dynamically \citep[][]{Alexander17,Mooley18a,Mooley18b,Mooley18c,Margutti18,Dobie18,Corsi18,Troja18,Alexander18,Resmi18,Ghirlanda19,Hajela19}. In addition to afterglow emission from a relativistic jet, a second synchrotron afterglow component caused by the dynamical/kilonova merger ejecta is predicted to dominate the radio lightcurve and to peak at low radio frequencies years following the merger \citep[][]{NakarPiran11,Hotokezaka16}. This never-before-seen radio afterglow would contain important information regarding the merger ejecta, which, for instance, can be used to make inferences about the elusive equation of state of nuclear matter. Apart from incoherent later-time radio emission, coherent emission produced around the merger time has long been predicted by various mechanisms \citep[e.g.][]{Hansen2001,UsovKatz,PshirkovPostnov,piro2012}. In particular, it has been suggested that a fraction of fast radio bursts (FRBs) could be accounted for by prompt emission arising from the coalescence of compact objects \citep[e.g.][]{totani13,FalckeRezzolla,Dorazio16}. Low-latency triggered radio observations covering the localization region of GW merger events would enable us to directly test these theories \citep{Chu16,RowlinsonAnderson20,Gourdji2020}.

The Low-Frequency Array \citep[LOFAR;][]{LOFAR13} has been used with its large field-of-view to follow-up on GW events at 145\,MHz since the first LIGO/Virgo observing run \citep[][]{Broderick15a,Broderick15b,AbbottLOFAR,Broderick17a,Rowlinson17a,Rowlinson17b}. In particular, the location of GW170817 was observed on several epochs, the results of which are reported in \citet{abbott17EM} and \citet{Broderick20}. The event's low elevation relative to LOFAR greatly affected the sensitivity achievable and thus the ability to place constraints on the radio light curve at low radio frequencies. LOFAR late-time follow-up results from the third GW observing run will be reported by Gourdji et al. (in preparation).

In this paper we present our strategy to search for GW merger radio afterglows in wide-field LOFAR data. We use wide-field multi-epoch LOFAR observations of G299232, a since ruled-out BH-NS GW trigger from the second GW observing run (O2), to establish and demonstrate our observing, calibration and transient search strategy, to guide future LOFAR follow-up of GW triggers and to examine the background of unrelated transient events.  In section \ref{sec:method}, we describe our observations, data calibration and imaging. In section \ref{sect:transient search}, we outline the three transient search methods undertaken. Our results are presented and discussed in section \ref{sect:results}. We discuss the prospects for LOFAR observations during the upcoming GW observing run in section \ref{sect:future} and conclude in section \ref{sect:conc}.
\setcounter{footnote}{-1}
\begin{figure}
    \centering
    \includegraphics[width=0.47\textwidth]{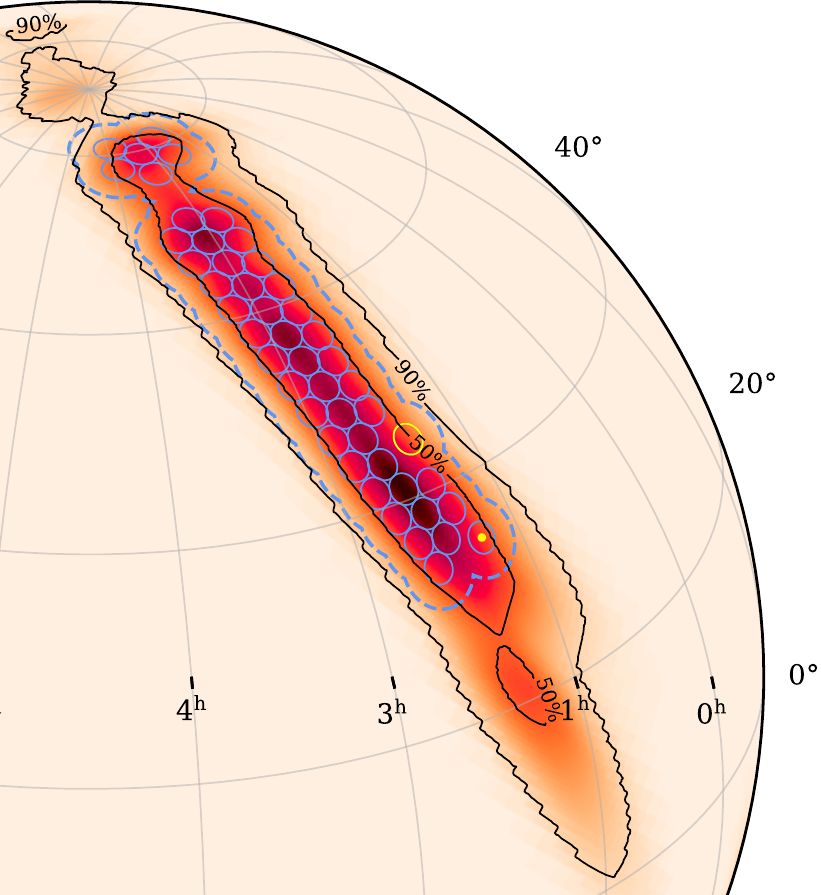}
    \caption{Part of the three-detector GW location probability density map of G299232 \citep[][]{g299232gcn}. Contours enclose 50 and 90 per cent of the probability density (though additional regions in the southern hemisphere are not included here). The full extent of the area searched in this analysis is enclosed by the blue dashed line, and corresponds to 47 overlapping LOFAR beams each searched out to a radius of 3.5\textdegree. The blue circles denote beam coverage of radius 1.4\textdegree, and the resulting unique sky area covered by these radii is 289.4\,deg$^2$. The optical transient detected contemporaneously with the GW trigger is shown by a yellow dot. The LOFAR beam placed at the centre of the neutrino candidate uncertainty region is shown in yellow. All beams are spaced by 2.8\textdegree, except for the two beams that were centred on the respective locations of the optical transient and the neutrino candidate. This figure was created using the \texttt{ligo.skymap} python module\protect\footnotemark.}
    \label{fig:skymap}
\end{figure}
 \footnotetext{{\url{https://lscsoft.docs.ligo.org/ligo.skymap}}}
\section{Observations and data reduction}
\label{sec:method}
\subsection{Observations}\label{sect:observations}
The data presented in this paper come from LOFAR follow-up observations of G299232 (also known as GW170825), a gravitational wave BH-NS merger candidate detected on 2017 August 25 at 13:13:37 UT during O2 and originally classified as EM-bright \citep[][]{g299232gcn}. The candidate was not detected in offline analysis by the LIGO/Virgo collaboration published two years later \citep{Abbott19O2}. The 50 and 90 per cent confidence intervals of the sky localization area were 451\,deg$^{2}$ and 2040\,deg$^{2}$, respectively. The GW location probability skymap and LOFAR coverage are depicted in Figure \ref{fig:skymap}. To probe the large localization region of the candidate event, we tiled 47 unique LOFAR beams each with a full width half maximum (FWHM) of 3.8\textdegree\ and using a beam spacing of around 2.8\textdegree\ (approximately equivalent to the beam FWHM divided by $\sqrt{2}$) to enclose $289.4$\,deg$^{2}$ of the localization region within beam radii of 1.4\textdegree\ (blue circles in Figure \ref{fig:skymap}), but with further coverage at the outer edges of the beam pattern (we search out to beam radii of 3.5\textdegree, represented by the dashed blue line in Figure \ref{fig:skymap}). We henceforth refer to the area of sky covered by a single LOFAR beam as a `sub-field'.  Each sub-field was observed for 225 minutes at 3 epochs corresponding to roughly 1~week, 1~month and 3~months following the GW trigger, as described in Table \ref{tab:obs}. A total of 48 LOFAR beams were used, but two were centred on the position of a neutrino candidate \citep[][]{GCNicecube,g299232Icecubeb} to provide an opportunity to stack the data and potentially improve sensitivity, leaving 47 unique sub-fields. The data were collected using LOFAR's HBA Dutch stations (24 core stations and 14 remote stations) in the \textsc{HBA\_DUAL\_INNER} configuration. Per epoch, we performed four 8-hour observing runs using two pointings.  To optimize the \textit{u-v} coverage, we alternated between two pointings every 25 minutes during each 8-hour observing run. Each pointing direction was comprised of six beams centred at 144\,MHz. Each beam, in turn, consisted of 81 frequency sub-bands with a bandwidth of 195.312\,kHz, yielding a total beam bandwidth of 15.82\,MHz. Each observation was book-ended by a ten minute calibrator scan of 3C48 and 3C196, in that order. The data were recorded using 2-second integration time-steps and 64 channels per sub-band. 

\subsection{Calibration and imaging}\label{sect:reduction}
Immediately following the observations, both the calibrator and target data were averaged in time and frequency to 10~seconds and 48.82\,kHz (4 channels per sub-band) during pre-processing to reduce data volume and Cas~A was subtracted from the visibilities using the so-called demixing procedure due to its close proximity. Standard LOFAR RFI flagging was also carried out during pre-processing, as described in \citet{Offringa12}.

The interferometric data were calibrated using \textsc{prefactor}\footnote{\url{https://github.com/lofar-astron/prefactor}}, a pipeline developed to correct for instrumental and ionospheric effects present in LOFAR datasets \citep{prefactor}. The \textsc{prefactor} calibration process begins by comparing observations of a calibrator source to the LOFAR skymodel of the calibrator source. This step provides direction-independent gain corrections that can be applied to the target observations. Once the target data are calibrated, they are imaged using the full bandwidth via \textsc{WSclean}\footnote{\url{https://gitlab.com/aroffringa/wsclean/}} \citep{wsclean14}, which is the final data product used for our scientific analysis.

We used 3C196 to calibrate our data using the Pandey flux density model \citep[which is consistent with the Scaife \& Heald model;][]{Pandey3C196}, except for observations taken on 2017 August 31, where station data were missing and so the back-up calibrator scans of 3C48 (\citealp{ScaifeHeald12} flux density model) were used instead. The sub-bands comprising each beam were concatenated into three groups of 27 sub-bands, each with a total bandwidth of 5.273\,MHz. The gain solutions from the calibrator were then applied to the target data. Sky models from the 147.5-MHz TIFR Giant Metrewave Radio Telescope Sky Survey First Alternative Data Release \citep[TGSS;][]{Intema17} were used for direction-independent phase corrections for our target observations.

The resulting calibrated data were used as input for \textsc{WSclean} to create image products. These include a primary-beam corrected image covering the full 15.82\,MHz observing bandwidth to be used for analysis, as well as individual images of the three groups of sub-bands (centered at 139\,MHZ, 144\,MHz, and 150\,MHz each with bandwidths of 5.273\,MHz) for reference purposes should a transient candidate appear in the full-bandwidth image. The \textsc{WSClean} options were optimally chosen for point sources and we used Briggs weighting with robustness parameter 0.5. We used baselines out to a maximum length of approximately 16\,km. The resulting images have a pixel resolution of 5\arcsec\ to sample the restoring beam of our images (2D Gaussian fit of the image point spread function), which on average has a major axis of 26.5\arcsec\ and minor axis of 18.2\arcsec.

\begin{figure}
    \centering
    \includegraphics[width=0.47\textwidth]{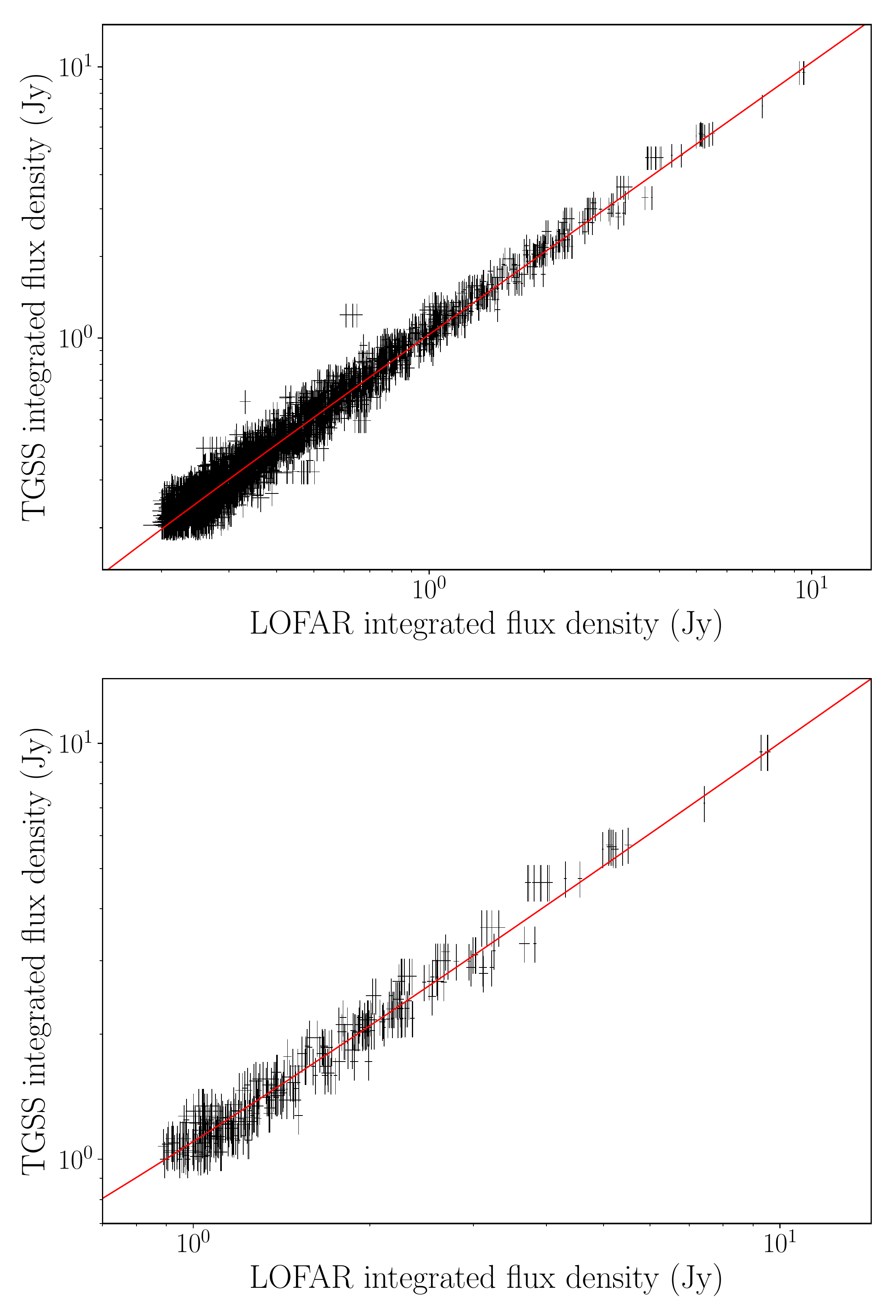}
    \caption{Comparison of source flux densities measured by TGSS and by \textsc{TraP} for the LOFAR data. Only compact sources with integrated flux densities exceeding 200\,mJy (top panel) and 1\,Jy (bottom panel) are considered. The fits were obtained via orthogonal distance regression and find a TGSS to \textsc{TraP} flux density ratio of 1.04 (top) and 0.99 (bottom). The \textsc{TraP} error bars correspond to the $1\sigma$ uncertainties on integrated source flux density values, and the TGSS error bars correspond to the same $1\sigma$ uncertainty plus an additional fractional error of 10 per cent, as reported in the TGSS catalogue.}
    \label{fig:flux_scale}
\end{figure}

We checked the flux density scale of our data by comparing to the TGSS catalogue (Figure \ref{fig:flux_scale}), as described in section \ref{meth:tgss}. We fit the data using orthogonal distance regression, which takes into account the uncertainties in both catalogues' source fluxes. We considered bright and compact matched sources with integrated flux density measurements in excess of 1\,Jy\,beam$^{-1}$ in the TGSS catalogue. For this, we used only those sources that were fit by a single Gaussian in the TGSS catalogue and that have an integrated flux to peak flux ratio less than 1.3 (the origin of this value is discussed in section \ref{meth:tgss}). Sources that are compact in the TGSS catalogue should generally be compact in our \textsc{TraP} catalogue as the major axis of the PSFs are comparable.  We find TGSS source flux measurements to have an average offset of 0.8 per cent. We attribute this negligible offset to the different flux models used (TGSS uses \citealp{ScaifeHeald12}) and possibly direction-dependent flux scale variations. For reference, we include a flux scale comparison using all compact TGSS sources above the 200\,mJy completeness threshold, in the top panel of Figure\,\ref{fig:flux_scale}.


\section{Transient search}
\label{sect:transient search}
For each individual sub-field, we process the set of three full-bandwidth images (one image per epoch) through the LOFAR transients pipeline (\textsc{TraP}, \citealp{Swinbank2015}), which identifies and extracts sources detected above $7\sigma$ in each image in a 3.5\textdegree\ radius from the beam centre. \textsc{TraP} identifies common sources between images and, if a source that is extracted in the second or third image is not matched to a source from an earlier image, it is flagged as a new source.  

We conduct three types of GW transient searches. In the first (section \ref{meth:cat}), we compare the catalogue of \textsc{TraP} sources found in our images to the TGSS catalogue that spatially overlaps. We compare to TGSS rather than to the LOFAR Two-metre Sky Survey \citep[LoTSS;][]{Shimwell19} as only a fraction of our field has been covered by LoTSS to date. This comparison serves additional functions apart from a general GW counterpart search: it allows us to verify the quality and flux scale of our catalogue, derive a transient false-alarm rate, and establish transient search criteria while retaining completeness. The second GW transient search (section \ref{meth:blind}) goes deeper in sensitivity and consists of comparing the \textsc{TraP} sources found between epochs and investigating potential new sources. The third search (section \ref{meth:target}) consists of targeted searches at the location of potential GW-related transient candidates detected at other wavelengths.


\subsection{Comparison to TGSS}
\label{meth:cat}
\subsubsection{Catalogue properties}
\begin{figure*}
    \begin{subfigure}[b]{0.49\textwidth}
        \centering
        \includegraphics[width=\textwidth]{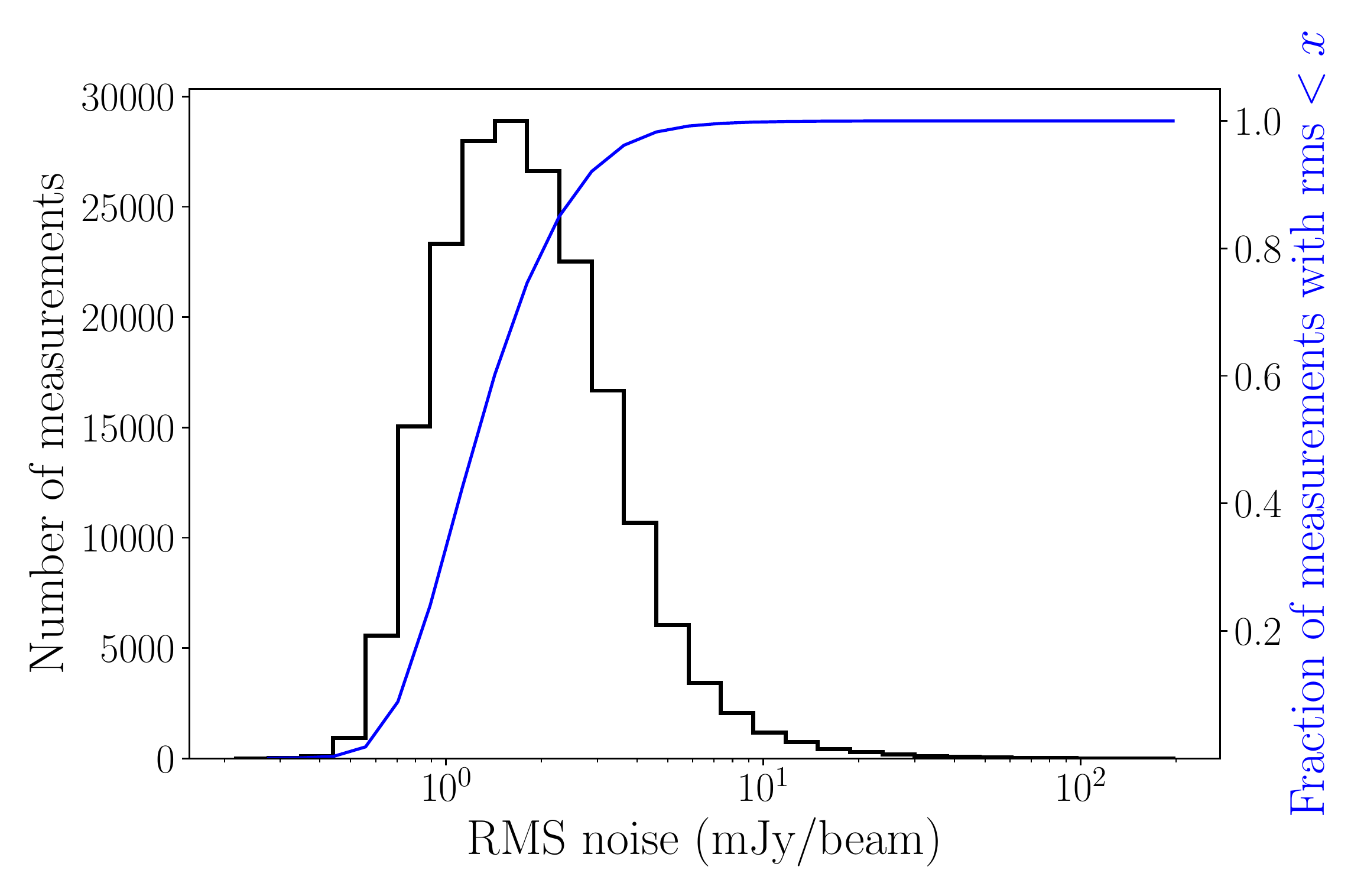}
        \caption{}
    \end{subfigure}
    \hfill
    \begin{subfigure}[b]{0.49\textwidth}
        \centering
    	\includegraphics[width=\textwidth]{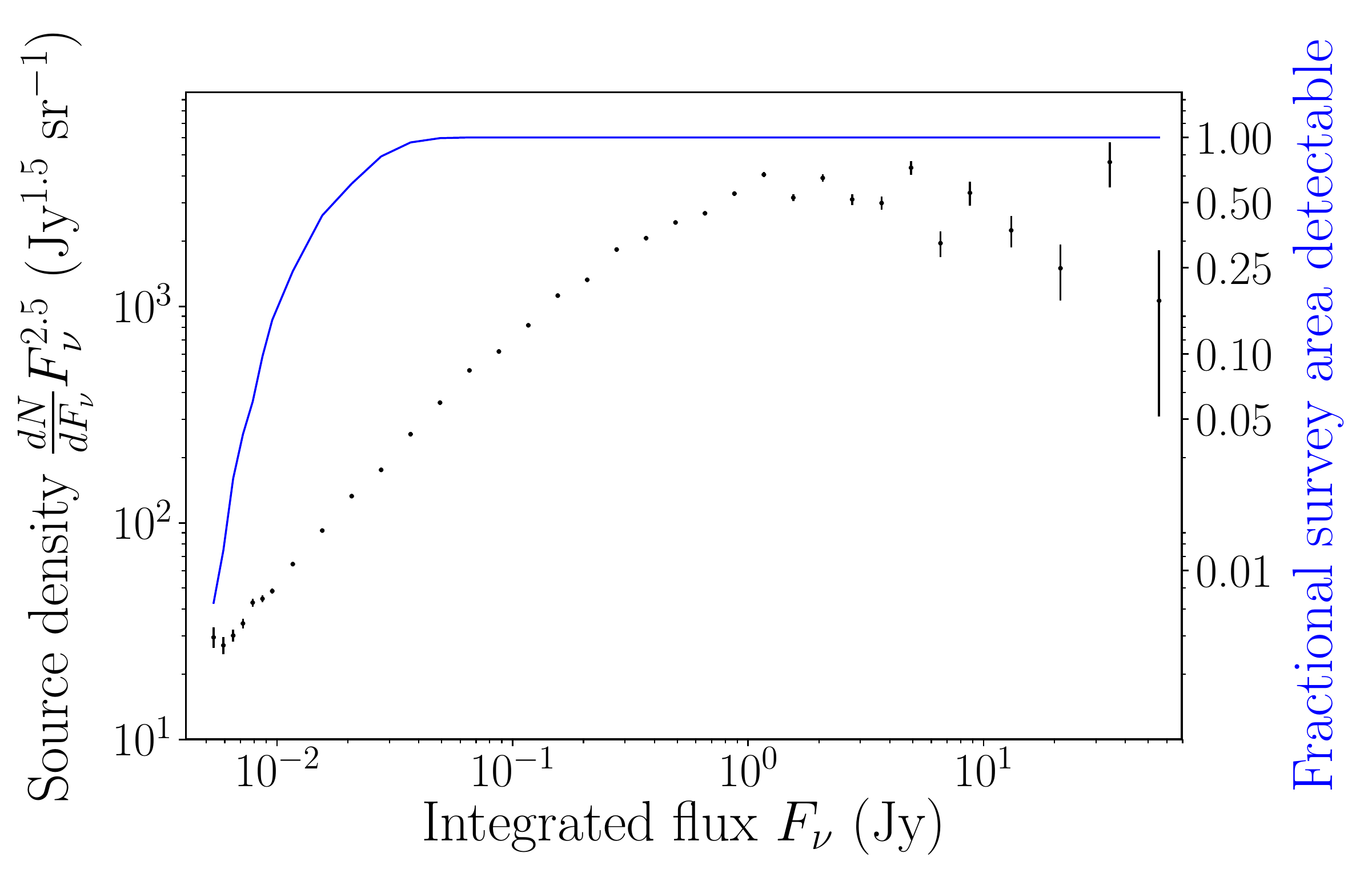}
        \caption{}
    \end{subfigure}
\caption{Left: distribution of nearly $200~000$ rms noise measurements taken across 141 LOFAR images. The blue curve represents the cumulative distribution of rms noise measurements. Right: Euclidean-normalized differential source density, with counts weighted by the fractional survey area detectable to correct for varying sensitivity between images as well as within a single image (as a function of distance from beam centre). All counts are normalized by the total survey area to obtain source density values. Error bars correspond to the propagated Poissonian error on each count. The data are tabulated in Table \ref{tab:sources}. The blue curve represents the inverse of the weights given in Table \ref{tab:sources} and corresponds to the fractional survey area in which a source with flux $F_{\nu}$ is detectable.  }
\label{fig:rms}
\end{figure*}

The TGSS survey and our data have comparable resolutions. The TGSS survey uses a 25$\arcsec$ circular restoring beam for the declinations observed in this study. We note that the emission we are targeting would appear as an unresolved point source in our data. First and foremost, we account for TGSS's lower sensitivity. TGSS is $100$ per cent complete above a flux of 200\,mJy. The detection threshold used for TGSS is $7\sigma$, where $\sigma$ is the local noise, the median of which is 3.5\,mJy\,beam$^{-1}$ \citep{Intema17}. Therefore, the median sensitivity of the survey is 24.5\,mJy. For comparison, the median local rms and sensitivity ($7\sigma$) of our LOFAR data is 1.7\,mJy and 12.0\,mJy respectively, with 80 per cent of our data having a sensitivity below 16\,mJy (see Figure \ref{fig:rms}). To account for differences in sensitivity between images as well as the decrease in sensitivity from the beam centre of a given image, we calculate the average rms $\sigma_\mu$ in four annuli of equal area for each image. We divide the entire range of fluxes $F_\nu$ of all the sources blindly detected by \textsc{TraP} into $N$ bins and weight the number of sources found in each bin by the total fractional area where the condition $F_\nu>7\sigma_\mu$ is met. This allows us to construct a true log($N$)-log($F_{\nu}$) source count distribution and compare the probability of source detection in our data (Figure \ref{fig:rms}). The results are tabulated in Table \ref{tab:sources}. The worst $7\sigma$ sensitivity of the complete dataset is 50\,mJy.

\subsubsection{Catalogue comparison and transient search}
\label{meth:tgss}
After setting a lower flux cut of 200\,mJy, we remove spurious \textsc{TraP} source detections caused by correlated noise or artefacts from sidelobes around bright sources (see Figure \ref{fig:cand types}). These can be identified rather easily as they are not expected to be consistently extracted in all epochs and typically have large positional uncertainties. We filter out spurious sources by excluding non-persistent sources (i.e. those not detected in all three LOFAR images) with positional error radii greater than $1''$. The value was determined heuristically during the catalogue comparison described below.

To compare source catalogues, we developed an algorithm that uses \textsc{astropy}'s \texttt{match\_to\_catalogue\_sky} tool \citep[][]{astropy:2013,astropy:2018}. This tool finds the closest source from a given reference catalogue to each source in a given input catalogue. The sky positions of each nearest association pair found by this tool are then used to calculate de Ruiter radii ($r_{ij}$), defined as the distance between source pairs ($ij$) weighted by their $1\sigma$ positional uncertainties:
\begin{equation}
    r_{ij} = \sqrt{\frac{\Delta\alpha_{ij}^2}{\sigma^2_{\alpha,ij}} + \frac{\Delta\delta{ij}^2}{\sigma^2_{\delta,ij}}
    } \,,
\end{equation}
where $\alpha$ is right ascension, $\delta$ is declination and $\sigma^2_{ij}=\sigma_i^2+\sigma_j^2$ calculated separately for both $\alpha$ and $\delta$.  The dimensionless de Ruiter radius calculated for each pair is used to determine whether the paired sources are truly associated. The probability density distribution of  positional differences  between catalogues of a given radio source due to measurement errors is the Rayleigh distribution \citep{deRuiter77}. This distribution can be used to give the probability, $p$, of having a source association at $r_{ij}\geq\rho$:
\begin{equation}
p(r_{ij}\geq\rho) = e^{-\rho^2/2}\,.
\end{equation}For this work, we use a threshold of $r_{ij}<5.68$, which corresponds to missing one in $10^{7}$ counterparts \citep{scheers11}. Repeat associations are filtered by finding the next nearest association (i.e. smallest de Ruiter radius) until only unique reference catalogue sources are associated. The mean source pair separation between catalogues is 1.4\arcsec.

There are two aspects that challenge a direct catalogue comparison. The first has to do with the different ways in which complex sources are extracted and modelled. TGSS models complex sources using multiple Gaussian fits, whereas \textsc{TraP} models all sources using a single Gaussian, and does not distinguish overlapping sources. Thus, comparing the catalogues directly will lead to false TGSS transients as, for a given complex source or cluster of sources, there will be multiple unique TGSS sources catalogued but only one \textsc{TraP} source. Given that the transients we are targeting will be unresolved point sources, we include only those TGSS sources that were fit with a single Gaussian (labeled `S' in the TGSS catalogue).

The second challenge is distinguishing between resolved and unresolved sources consistently between catalogues. While an unresolved point source should theoretically have a ratio between integrated flux and peak flux ($\frac{F_{\textnormal{tot}}}{ F_{\textnormal{peak}}}$) equal to one and source size fit equal to the restoring beam, this is often not the case due to imperfect calibration and finite visibility sampling. These effects naturally differ between catalogues resulting in different degrees of smearing, which causes this ratio ($\frac{F_{\textnormal{tot}}}{ F_{\textnormal{peak}}}$, henceforth referred to as the `compactness') to be greater than one for a given source. Furthermore, the amount of smearing is strongly dependent on the signal-to-noise ratio (S/N). Thus, a universal cut on compactness is not possible a priori. Fortunately, \citet{Intema17} provide an empirical distinction between resolved and unresolved sources as a function of S/N in the TGSS catalogue. We use Figure 11 from \citet{Intema17} to conservatively choose a large value that encompasses all unresolved sources above 200\,mJy ($\frac{F_{\textnormal{tot}}}{F_{\textnormal{peak}}}<1.3$). This leaves 3936 (not all unique) TGSS sources to match to our \textsc{TraP} catalogue. As we do not yet know what the corresponding compactness threshold is in our data, we first look for matches to this subset of TGSS sources in our \textsc{TraP} catalogue (the mean source property values are used for unique sources that are detected across multiple LOFAR epochs), which we limit to sources with $\frac{F_{\textnormal{tot}}}{ F_{\textnormal{peak}}}<5$ and $F_{\textnormal{tot}}>200$\,mJy as a first pass. This results in 8572 (not all unique) \textsc{TraP} sources to match from. A small fraction of TGSS sources with $F_{\textnormal{tot}}$ slightly greater than $200$\,mJy will be unmatched if their \textsc{TraP} counterpart has $F_{\textnormal{tot}}$ slightly less than this cut. There are 197 such cases and matches are recovered by searching below the lower flux cut. Five TGSS sources remain unassociated.

We then proceed to look for unresolved \textsc{TraP} sources without a counterpart in the TGSS catalogue, as these would be transient candidates. To determine a boundary between unresolved and resolved sources in our data above 200\,mJy, we take the maximum value in the distribution of compactness values of the \textsc{TraP} matches. This process is done for each sub-field, as the compactness boundary between resolved and unresolved sources varies between sub-fields (the mean compactness boundary is 1.7, with the largest value at 2.3). These cuts are very conservative given that the set of TGSS sources matched from is certain to contain some resolved sources, and so likewise our derived set of \textsc{TraP} unresolved sources is certain to include some resolved sources. We are left with 2324 unmatched \textsc{TraP} sources after applying this compactness filter.

A TGSS counterpart is not immediately found for a significant fraction of these \textsc{TraP} sources because, for reasons not entirely clear, some compact sources are classified as complex sources (i.e. not type `S') in the TGSS catalogue. This likely has to do with subtleties in the source extraction step and we note that \textsc{TraP} uses a different source extraction tool \citep[\textsc{PySE;}][]{pyse} than is used for TGSS \citep[\textsc{PyBDSF;}][]{pybdsf}. Therefore, for the remaining unmatched sources, we search for counterparts in the full catalogue of TGSS sources (that is, allowing all fit-types, not only single-Gaussian sources and no compactness or lower flux cuts) and successfully find counterparts. TGSS matches are found for all 2324 remaining \textsc{TraP} sources.

\subsection{Deep blind transient search}
\label{meth:blind}
The sensitivity of the previous transient search is limited by the completeness of the TGSS catalogue. In order to go deeper, we look for transient sources between our images. Low-frequency radio waves from a GW afterglow caused by a jet are expected to be visible on month time-scales. The set of observations from our first epoch (taken one week after the GW trigger) should not contain a counterpart and therefore provide reference images that allow us to establish whether a new source has appeared in a later epoch. Depending on the viewing angle of the jet, emission may first appear in the second or third epoch. Thus, we search for new sources that \textsc{TraP} detects for the first time in the second or third LOFAR image. Searching down to lower fluxes, however, naturally increases the number of spurious sources, and criteria to separate false positives from true detections become muddled. We find that these new sources can generally be classified into five categories (examples are provided in Figure \ref{fig:cand types}):
\begin{enumerate}
\item  extended sources with fit parameters that differ significantly between images, which prevent them from being associated to one another by \textsc{TraP};
\item  imaging artefacts from sidelobes around bright sources;
\item noise;
\item  faint sources detected around the detection threshold in the second and/or third image, but that were undetected in at least the first image due to higher local rms;
\item potential astrophysical transient.
\end{enumerate}

As in the previous section, we are faced with the difficulty of distinguishing between resolved and unresolved sources; the cuts found in the previous section pertain to sources above 200\,mJy and are not applicable for fainter detections. Furthermore, we find that sources close to the detection threshold can have fits that include surrounding correlated noise, which can increase the fit size and hence the compactness value. Therefore, we conservatively filter out sources with compactness values exceeding 5. This criterion also manages to filter out some unwanted type \textit{ii)} candidates as well as the small fraction of type \textit{iii)} candidates caused by correlated noise, as these often have large fit sizes.

Due to the larger fit sizes of faint sources, the positional errors of these sources will be larger and the previously used criterion of excluding sources with error radii larger than 1\arcsec\ is too stringent. Therefore a different method is required to exclude the large number of spurious type \textit{ii)} candidates. We start by creating a catalogue of \textsc{TraP} sources that were blindly detected in all epochs above $60 \sigma$. The remaining compact transient candidates are compared to this catalogue of persistent and bright sources, and are excluded if they fall within a distance of 100\arcsec\ (roughly 4 times the size of the main lobe of the restoring beam and includes the brightest sidelobes) of one of the sources in the catalogue. In order to avoid filtering out bright compact transient candidates, we include all sources above $15 \sigma$, regardless of their proximity to a bright persistent source. These threshold values were determined heuristically and in a way that strikes a balance between percentage of the sub-field excised and number of category \textit{ii)} candidates that still remain. The chosen thresholds correspond to excising about one per cent of the total amount of (overlapping) sky probed and we find them to work reasonably well across all sub-fields.

Most of the remaining candidates are category \textit{iv)} sources and a fraction of \textit{ii)} artefacts that passed through the filter likely because they originate from persistent sources below our $60\sigma$ threshold. The positions of the remaining transient candidates are fed through \textsc{TraP} again, to force a flux measurement at these positions in the full-bandwidth image of each epoch using the respective restoring beams. This effectively assumes that all new source candidates are unresolved point sources. 

Finally, we identify candidates that are not present in the first epoch by checking whether the forced fits are consistent with forced fits of noise. To characterize a forced fit of genuine noise, we used \textsc{TraP} to perform forced extractions at random coordinates in an image, excluding those that happened to fall within 70\arcsec\ of a known source. The results from this analysis were used to establish a criterion to reject candidates. In particular, we compare the integrated flux ($F_\nu$) measurement to its $1\sigma$ error ($F_{\textnormal{err}}$) and keep candidates where $F_{\textnormal{err}}/F_\nu > 0.5$ or where $F_\nu<0$ in the first epoch (consistent with noise and no source being present). In contrast, we find that the fractional error is always less than 30 per cent for all persistent sources blindly extracted by \textsc{TraP} in all three epochs. These filters yielded 339 transient candidates to inspect manually.

\subsection{Targeted multiwavelength and multi-messenger counterpart search}
\label{meth:target}
In addition to a blind search, we can perform targeted searches at the locations of transient candidates detected at other wavelengths that could be related to the GW merger candidate. Within this large field, there were two transient candidates reported following the GW trigger. One was an IceCube Neutrino Observatory candidate detected 233 seconds prior to the GW event, with energy 0.39\,TeV \citep{GCNicecube}. For this reason, we centred two of our 48 LOFAR beams on the coordinates of the neutrino candidate and our search radius covers 85 per cent of the neutrino candidate's uncertainty region. Ultimately, for each epoch the higher-quality image was used for analysis. The second field of interest was a rapidly fading \textit{Swift}-UVOT optical transient detected two days after the GW trigger \citep{EmeryGCN,JonkerGCN}, on which we centred a LOFAR beam. 
The precisely known location of the optical transient enabled us to perform a targeted search for related radio emission. We used \textsc{TraP} to perform a forced fit matching the restoring beam at that location to measure the peak flux density and place upper limits on the radio emission (see section \ref{sect:forced}).


\section{Results and discussion}
\label{sect:results}
\subsection{Transient false negative rate}
The locations of the five TGSS sources that did not have a LOFAR counterpart from \textsc{TraP} were visually inspected in the corresponding LOFAR images. In all cases, the TGSS source is visually present and was rejected by our source association algorithm because the compactness value was larger than the cut imposed. These are rare instances where the TGSS source has one or more nearby sources that \textsc{TraP} does not distinguish. Here, \textsc{TraP} extracts the multiple sources as a single extended source with compactness greater than 5. 

We can use the TGSS sources missed by \textsc{TraP} in our catalogue comparison to determine the likelihood that a transient is missed by \textsc{TraP}, above 200\,mJy. For 1809 deg$^{2}$ of sky surveyed (there are 47 sub-fields, each with an area corresponding to $3.5^2\pi$\,deg$^2$, where 3.5 is the search radius of each sub-field in degrees), 5 out of 3936 TGSS sources were missed by \textsc{TraP}, yielding a false negative rate of 0.1 per cent. We note that this corresponds to the total overlapping sky area sampled and so not to unique sources. 

\subsection{Blind GW transient search}
\label{sec:disc blind}
The 339 transient candidates that remained after the candidate filter step were manually inspected. The three sub-band images were also considered during inspection as an astrophysical transient is expected to be visible across the full frequency band whereas spurious sources are typically narrow-band. We find that 301 transient candidates (91 per cent) were false positives caused by sidelobes and correlated noise. Thirty one candidates were visible in all three sub-band images and warranted further consideration.

\begin{figure}
    \centering
    \includegraphics[width=0.47\textwidth]{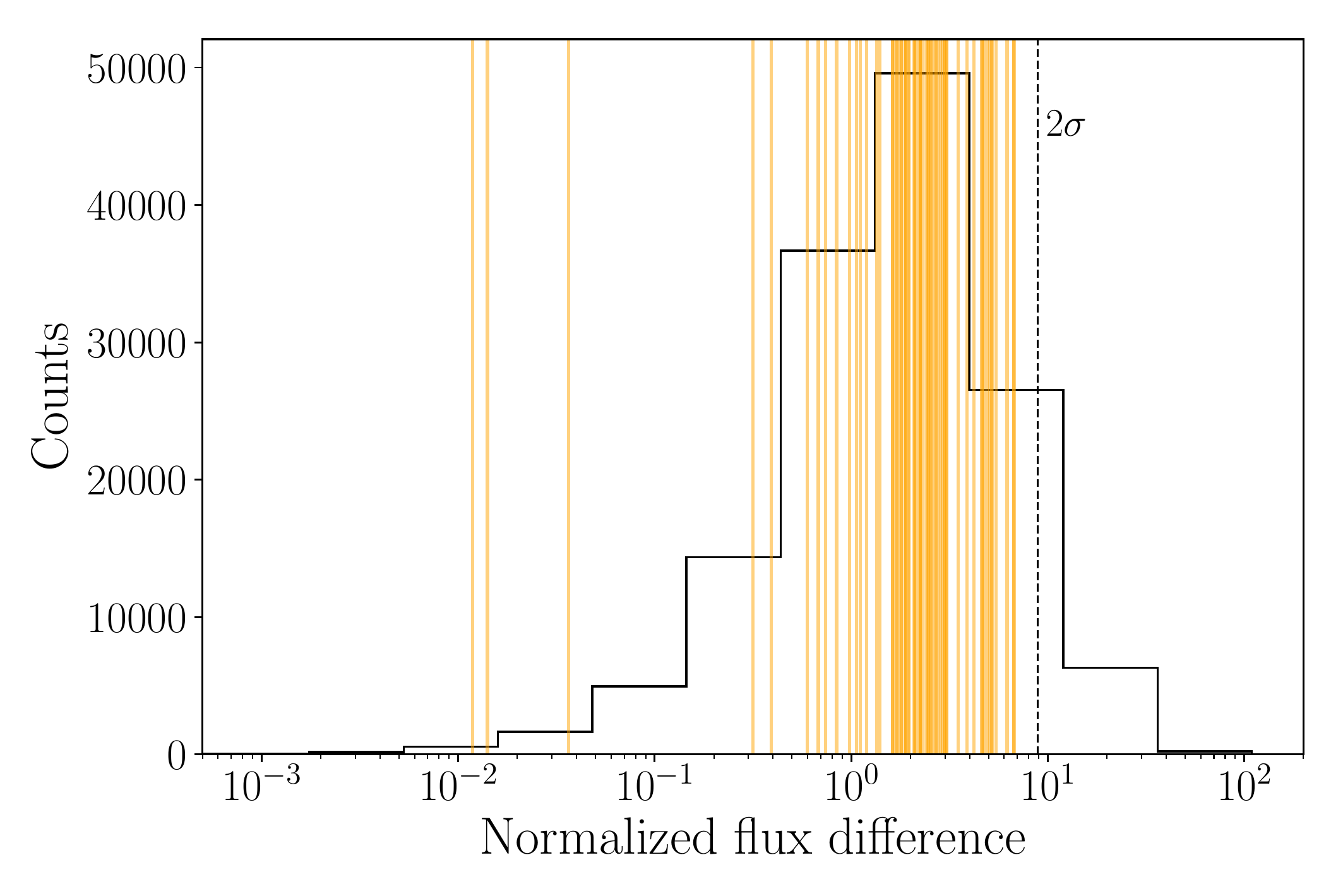}
    \caption{Distribution of source flux difference between images for all persistent sources and normalized by the uncertainties on each flux measurement summed in quadrature. The vertical yellow lines represent the normalized flux difference for 31 transient candidates that passed inspection, and are consistent with the overall source distribution. For reference, a vertical dashed line denoting the $2\sigma$ value of the distribution of persistent sources is included.}
    \label{fig:flux_diff}
\end{figure}

In order to determine whether a transient candidate's apparent relative increase in flux is significant, we investigate the distribution of peak flux differences between all persistent sources blindly detected by \textsc{TraP} in all sets of images with compactness values less than 5 (see section \ref{meth:tgss}). To quantify this, we calculate the absolute differences between peak flux measurements for a given source and normalize by the uncertainties ($\sigma_F$) summed in quadrature:
\begin{equation}
    \frac{|F_{\textnormal{peak},i}-F_{\textnormal{peak},j}|}{\sqrt{\sigma^2_{F,i}+\sigma^2_{F,j}}}\,,
\end{equation}
where $i$ and $j$ denote source measurements at different epochs (all three possible permutations are considered). We follow the same procedure for the transient candidates. The results of this analysis are shown in Figure \ref{fig:flux_diff} and demonstrate that the flux difference found between transient candidate images are insignificant. Specifically, the flux difference values of all transient candidates lie within $2\sigma$ of the distribution of flux differences of persistent sources.  Thus, we conclude that the fluxes of the transient candidates do not deviate significantly from the behaviour of the persistent sources and therefore no new transients are present in our data.  An assessment of general source flux variability is beyond the scope of this paper.

\begin{figure*}
    \centering
    \includegraphics[width=0.9\textwidth]{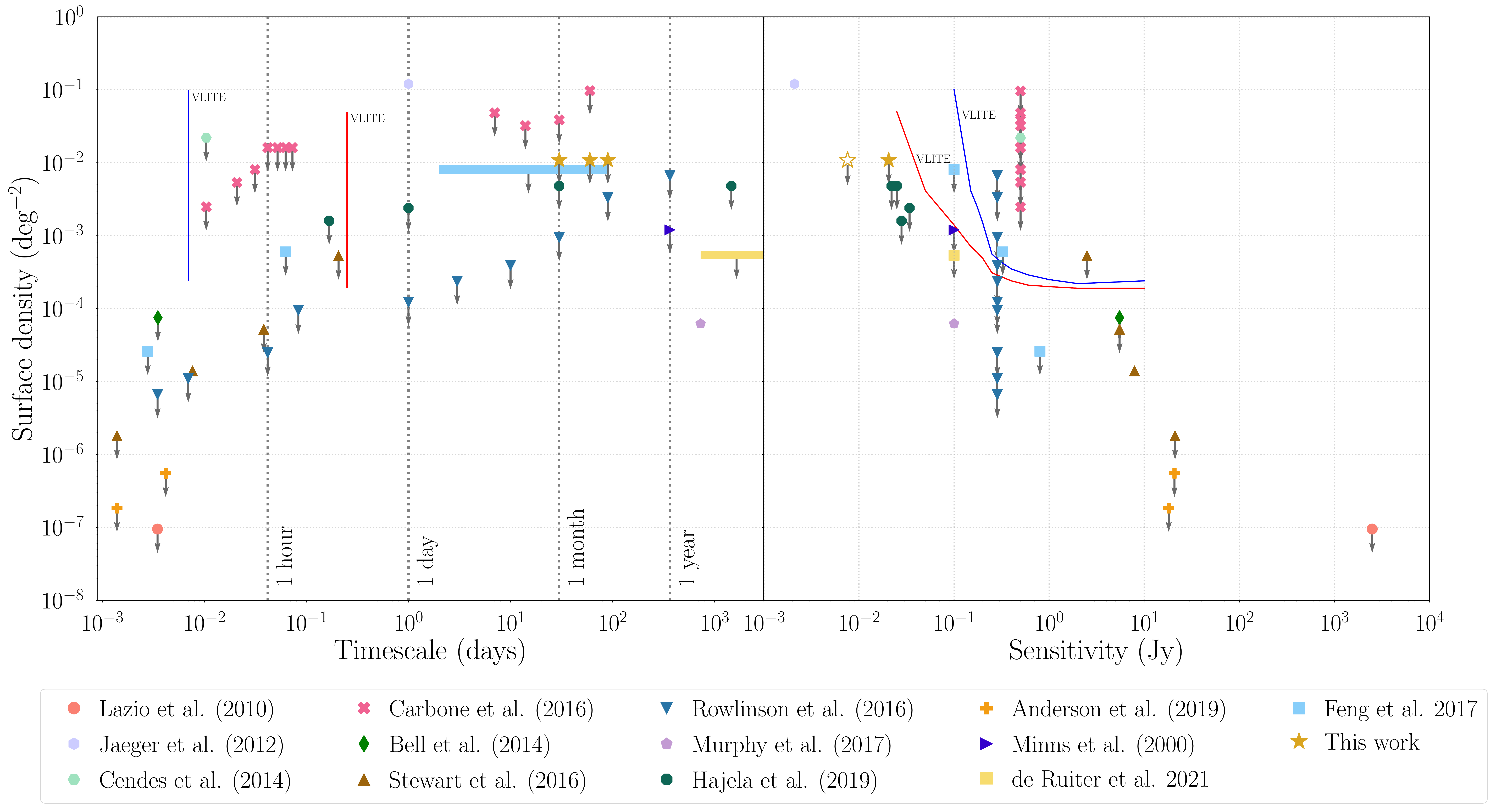}
    \caption{Overview of transient surface densities between about 60\,MHz and 340\,MHz at timescales greater than one minute. The left panel shows the transient surface density as a function of survey time-scale probed, while the right panel shows the surface density as a function of sensitivity probed. The down-facing arrows represent upper limits, whereas data points with no down-facing arrows represent actual transient detections. The limits set by this study are represented by gold stars. A hollow star depicts the median sensitivity (8 mJy) across the 277 deg$^2$ of field used for this analysis (see section \ref{sec:disc blind}). This is included in order to compare to \citet{Hajela19}, the next most sensitive survey on month timescales, which only quotes the median sensitivity. The blue and red lines correspond to limits set by the VLITE survey from \citet{Polisensky16}. A running catalogue of slow radio transient surveys can be found in \citet{Mooley16}. Galactic centre radio transients have been excluded from this figure.}
    \label{fig:transients}
\end{figure*}

Our relatively high-sensitivity data of a large field taken at 144\,MHz at these three particular epochs presents an opportunity to probe a hitherto unexplored area of transient surface density phase space (Figure \ref{fig:transients}). To obtain transient surface density limits, we consider the amount of unique sky sampled (corresponding to the inner 1.4\textdegree\ radius of the LOFAR beams). We calculate the mean rms within this radius for each image. We exclude 2 sub-fields (with right ascensions corresponding to 25.04\textdegree\ and 35.45\textdegree, see Table \ref{tab:obs}) that contained outliers in the image rms comparison. Conservatively taking the maximum mean rms value and multiplying by our detection threshold, we obtain a $7\sigma$ survey sensitivity of 20\,mJy for 277\,deg$^2$ of sky coverage. We follow \citet{Rowlinson16} to calculate an upper limit on the transient surface density at the $95$ per cent confidence level and find 0.01\,deg$^{-2}$ for time-scales corresponding to about one, two and three months. As can be seen in Figure \ref{fig:transients}, this analysis is the most sensitive transient search at low radio frequencies on those time-scales: there are only three other surveys at comparable frequency and time-scales and our data are at least three times more sensitive.

\subsection{Targeted search limits}
\label{sect:forced}
No transient candidates were found in the neutrino candidate's error region.The results of forced flux measurements at the location of the UVOT optical transient are summarized in Table \ref{table:uvot}. Significant radio emission is not detected in any of the three images. That the beam was centred on the position of the optical transient enables us achieve 3$\sigma$ upper limits down to 1.5\,mJy\,beam$^{-1}$ without any direction-dependent calibration. Note that this sub-field unfortunately happens to contain the poorest-quality images of all sub-fields.
 
\begin{table}
\centering
\begin{tabular}{ccc}
 & $3\sigma$ limit & Peak flux density\\
Epoch & (mJy beam$^{-1}$) & (mJy beam$^{-1}$)  \\\hline
1     & 1.5                                                       & $1.0 \pm 0.9$                \\
2     & 1.8                                                       & $2 \pm 1$                    \\
3     & 4.9                                                       & $4 \pm 2$                   
\end{tabular}
\caption{Forced measurements at the location of the UVOT optical transient. The location corresponds to the centre of a LOFAR beam. The $3\sigma$ limits correspond to 3 times the local rms value as measured by \textsc{TraP}.}
\label{table:uvot}
\end{table}

\section{Prospects for future LOFAR GW counterpart searches}
\label{sect:future}
The results presented in this paper demonstrate LOFAR's ability to provide deep limits over large fields, making it a unique and powerful tool to cover vast swaths of poorly localized gravitational wave merger events. In the next GW observing run, the location probability fields are projected to be significantly smaller: the four-detector GW network forecasts median 90 per cent credible regions for the localization area of BNS and BH-NS mergers of $33\pm5$\,deg$^2$ and $50\pm8$\,deg$^2$, respectively \citep[][]{abbott2020prospects}. A single LOFAR pointing consisting of 7 beams in a hexagonal pattern will cover roughly 40\,deg$^2$ at optimum sensitivity (1.4\textdegree\ radius around beam centres) with further coverage away from the centre of the beam pattern. The median sensitivity achieved in this analysis using a 1.4\textdegree\ search radius is 8\,mJy. These much smaller field sizes will enable us to reduce the data more carefully and increase data quality. In particular, direction-dependent calibration (DDC), which is computationally expensive and unfeasible for fields as large as the one analyzed in this work, can be performed. We have purposefully calibrated our data in a way that is compatible with further reduction via \textsc{KillMS} \citep[][]{Tasse14a,Tasse14b,SmirnovTasse15} and \textsc{DDFacet} \citep[][]{Tasse18} or \textsc{Factor}\footnote{\url{https://github.com/lofar-astron/factor}} for DDC. In particular, \textsc{KillMS} and \textsc{DDFacet}\footnote{\url{https://github.com/mhardcastle/ddf-pipeline}} are used by the LoTSS surveys team who regularly achieve 70\,$\upmu$Jy\,beam$^{-1}$ median rms noise levels for their 8-hour integration images \citep[][]{Shimwell19}. DDC would improve sensitivity and resolution, and should significantly reduce \textsc{TraP} transient false positives by minimizing sidelobe artefacts, which will allow us to relax our search filters and throw a smaller fraction of the sky away in order to decrease the number of candidates. Furthermore, a mosaic of the field can be constructed before searching for transients, to obtain deeper and more uniform sensitivity across the search area. Finally, in cases where an EM counterpart has already been identified, all LOFAR sub-bands can be used to place a single LOFAR beam at its location, thus increasing sensitivity to LoTSS levels (i.e. $3\sigma$ sensitivity of $210$\,$\upmu$Jy with DDC). An analysis of LOFAR data corresponding to GW triggers detected in O3 is in preparation and will showcase some of the aforementioned improvements and scenarios (Gourdji et al. in prep.).

For GW candidate fields that LoTSS has previously surveyed, we can use LoTSS data as our reference image and compare our later-time LOFAR observations to that catalogue of sources directly to search for GW transients and to perform data quality control (as opposed to performing our own reference observation within a week of the merger and then comparing to TGSS). In that case, the sensitivity of the search described in section \ref{meth:tgss} would be limited by the LOFAR data, rather than by the TGSS catalogue's much higher 200\,mJy completeness threshold. In addition, many of the catalogue comparison challenges discussed in section \ref{meth:tgss} would be diminished or gone altogether.

\citet{Broderick20} explored the feasibility of using LOFAR to detect the afterglow of GW170817, had the event occurred further north on the sky. They showed that the light-curve, extrapolated to 144\,MHz, peaks at about $500$\,$\upmu$Jy, thus necessitating DDC. They also considered the scenario where an event similar to GW170817 takes place at a distance of 100\,Mpc and in a denser (factor 10) interstellar medium and show that, for various afterglow models, the LOFAR sensitivity would have to be at least 1\,mJy, which again requires DDC. The brightness distribution of GW merger afterglows is, however, highly uncertain given the very small sample size and considerable model uncertainties.

Finally, LOFAR can be used to rapidly observe GW triggers within 5 minutes, to search for related prompt emission. This strategy has been successfully tested with triggered LOFAR observations of gamma-ray bursts \citep[][]{Rowlinson19,Rowlinson20}. LOFAR will be one of few radio facilities capable of probing coherent radio emission at such short time-scales post-merger, and particularly at such sensitivities. For instance, the Murchison Widefield Array (MWA) can be on source within approximately 20 seconds but is less sensitive than LOFAR by at least an order of magnitude \citep[][]{Anderson21}. It is as yet unclear whether prompt coherent emission is produced and whether it would be able to escape the merger environment. Nonetheless, a detection would immediately identify the precise location of the merger and would undeniably link compact mergers as a progenitor of FRBs. The recent detection of FRBs with LOFAR shows that they can at least be detected at low frequencies \citep{pleunis21,pastor21}.

\section{Conclusions}
\label{sect:conc}
We have demonstrated a strategy to search for slow transients related to gravitational wave merger events in wide-field LOFAR data. In particular, we have presented three different transient search methods involving the comparison to an existing catalogue of radio sources, a blind search between multi-epoch LOFAR images and a targeted search at the location of transients detected at other wavelengths. These methods are telescope agnostic and can be applied to other interferometric radio observations of GW events. We have not found any transients and place the most sensitive transient surface density limits at low radio frequencies to date on time-scales of order one month.

We have discussed LOFAR's prospects for the upcoming GW observing run, and show that LOFAR will be able to probe the full projected median localization area for GW merger events with a single 4-hour multi-beam observation down to a median sensitivity of 8\,mJy, before direction-dependent calibration. Thus, LOFAR will provide the deepest wide-field datasets probing the afterglow of GW merger events at low radio frequencies.
\section*{Acknowledgements}
We thank Mark Kuiack and Iris De Ruiter for helpful discussions. We thank the ASTRON Radio Observatory for setting up the observations used in this paper, and for preprocessing the data.

This paper is based on data obtained with the International LOFAR Telescope (ILT) under project code LC8\_005. LOFAR \citep{LOFAR13} is the Low Frequency Array designed and constructed by ASTRON. It has observing, data processing, and data storage facilities in several countries, that are owned by various parties (each with their own funding sources), and that are collectively operated by the ILT foundation under a joint scientific policy. The ILT resources have benefited from the following recent major funding sources: CNRS-INSU, Observatoire de Paris and Universit\'e d'Orl\'eans, France; BMBF, MIWF-NRW, MPG, Germany; Science Foundation Ireland (SFI), Department of Business, Enterprise and Innovation (DBEI), Ireland; NWO, The Netherlands; The Science and Technology Facilities Council, UK; Ministry of Science and Higher Education, Poland.

\section*{Data availability}
The LOFAR visibility data are publicly available at the LOFAR Long Term Archive (\url{https://lta.lofar.eu/}) under project code LC8\_005. The derived images used for this article's analysis will be shared on reasonable request to the corresponding author. A basic reproduction package (\url{10.5281/zenodo.5599729}) outlines the data reduction software and settings used in this analysis. It also provides the scripts and data required to reproduce the figures and tables of this paper.


\nocite{deRuiter21,Anderson19,Murphy2017,Feng17,Stewart16,Carbone16,Bell14,Cendes14,Lazio10,Jaeger12,Hyman2009,MinnsRiley2000}

\bibliographystyle{mnras}
\bibliography{main} 

\begin{thebibliography}{}
\makeatletter
\relax
\def\mn@urlcharsother{\let\do\@makeother \do\$\do\&\do\#\do\^\do\_\do\%\do\~}
\def\mn@doi{\begingroup\mn@urlcharsother \@ifnextchar [ {\mn@doi@}
  {\mn@doi@[]}}
\def\mn@doi@[#1]#2{\def\@tempa{#1}\ifx\@tempa\@empty \href
  {http://dx.doi.org/#2} {doi:#2}\else \href {http://dx.doi.org/#2} {#1}\fi
  \endgroup}
\def\mn@eprint#1#2{\mn@eprint@#1:#2::\@nil}
\def\mn@eprint@arXiv#1{\href {http://arxiv.org/abs/#1} {{\tt arXiv:#1}}}
\def\mn@eprint@dblp#1{\href {http://dblp.uni-trier.de/rec/bibtex/#1.xml}
  {dblp:#1}}
\def\mn@eprint@#1:#2:#3:#4\@nil{\def\@tempa {#1}\def\@tempb {#2}\def\@tempc
  {#3}\ifx \@tempc \@empty \let \@tempc \@tempb \let \@tempb \@tempa \fi \ifx
  \@tempb \@empty \def\@tempb {arXiv}\fi \@ifundefined
  {mn@eprint@\@tempb}{\@tempb:\@tempc}{\expandafter \expandafter \csname
  mn@eprint@\@tempb\endcsname \expandafter{\@tempc}}}

\bibitem[\protect\citeauthoryear{{Abbott} et~al.,}{{Abbott}
  et~al.}{2016}]{AbbottLOFAR}
{Abbott} B.~P.,  et~al., 2016, \mn@doi [\apjl] {10.3847/2041-8205/826/1/L13},
  \href {https://ui.adsabs.harvard.edu/abs/2016ApJ...826L..13A} {826, L13}

\bibitem[\protect\citeauthoryear{{Abbott} et~al.,}{{Abbott}
  et~al.}{2017a}]{abbott17GW}
{Abbott} B.~P.,  et~al., 2017a, \mn@doi [\prl]
  {10.1103/PhysRevLett.119.161101}, \href
  {https://ui.adsabs.harvard.edu/abs/2017PhRvL.119p1101A} {119, 161101}

\bibitem[\protect\citeauthoryear{{Abbott} et~al.,}{{Abbott}
  et~al.}{2017b}]{abbott17H0}
{Abbott} B.~P.,  et~al., 2017b, \mn@doi [\nat] {10.1038/nature24471}, \href
  {https://ui.adsabs.harvard.edu/abs/2017Natur.551...85A} {551, 85}

\bibitem[\protect\citeauthoryear{{Abbott} et~al.,}{{Abbott}
  et~al.}{2017c}]{abbott17EM}
{Abbott} B.~P.,  et~al., 2017c, \mn@doi [\apjl] {10.3847/2041-8213/aa91c9},
  \href {https://ui.adsabs.harvard.edu/abs/2017ApJ...848L..12A} {848, L12}

\bibitem[\protect\citeauthoryear{{Abbott} et~al.,}{{Abbott}
  et~al.}{2019}]{Abbott19O2}
{Abbott} B.~P.,  et~al., 2019, \mn@doi [\apj] {10.3847/1538-4357/ab0e8f}, \href
  {https://ui.adsabs.harvard.edu/abs/2019ApJ...875..161A} {875, 161}

\bibitem[\protect\citeauthoryear{{Abbott} et~al.,}{{Abbott}
  et~al.}{2020}]{abbott2020prospects}
{Abbott} B.~P.,  et~al., 2020, \mn@doi [Living Reviews in Relativity]
  {10.1007/s41114-020-00026-9}, \href
  {https://ui.adsabs.harvard.edu/abs/2020LRR....23....3A} {23, 3}

\bibitem[\protect\citeauthoryear{{Alexander} et~al.,}{{Alexander}
  et~al.}{2017}]{Alexander17}
{Alexander} K.~D.,  et~al., 2017, \mn@doi [\apjl] {10.3847/2041-8213/aa905d},
  \href {https://ui.adsabs.harvard.edu/abs/2017ApJ...848L..21A} {848, L21}

\bibitem[\protect\citeauthoryear{{Alexander} et~al.,}{{Alexander}
  et~al.}{2018}]{Alexander18}
{Alexander} K.~D.,  et~al., 2018, \mn@doi [\apjl] {10.3847/2041-8213/aad637},
  \href {https://ui.adsabs.harvard.edu/abs/2018ApJ...863L..18A} {863, L18}

\bibitem[\protect\citeauthoryear{{Alexander} et~al.,}{{Alexander}
  et~al.}{2021}]{Alexander2021}
{Alexander} K.~D.,  et~al., 2021, arXiv e-prints, \href
  {https://ui.adsabs.harvard.edu/abs/2021arXiv210208957A} {p. arXiv:2102.08957}

\bibitem[\protect\citeauthoryear{{Anderson} et~al.,}{{Anderson}
  et~al.}{2019}]{Anderson19}
{Anderson} M.~M.,  et~al., 2019, \mn@doi [\apj] {10.3847/1538-4357/ab4f87},
  \href {https://ui.adsabs.harvard.edu/abs/2019ApJ...886..123A} {886, 123}

\bibitem[\protect\citeauthoryear{{Anderson} et~al.,}{{Anderson}
  et~al.}{2021}]{Anderson21}
{Anderson} G.~E.,  et~al., 2021, arXiv e-prints, \href
  {https://ui.adsabs.harvard.edu/abs/2021arXiv210414758A} {p. arXiv:2104.14758}

\bibitem[\protect\citeauthoryear{{Andreoni} et~al.,}{{Andreoni}
  et~al.}{2020}]{Andreoni2020}
{Andreoni} I.,  et~al., 2020, \mn@doi [\apj] {10.3847/1538-4357/ab6a1b}, \href
  {https://ui.adsabs.harvard.edu/abs/2020ApJ...890..131A} {890, 131}

\bibitem[\protect\citeauthoryear{{Antier} et~al.,}{{Antier}
  et~al.}{2020}]{Antier2020}
{Antier} S.,  et~al., 2020, \mn@doi [\mnras] {10.1093/mnras/staa1846}, \href
  {https://ui.adsabs.harvard.edu/abs/2020MNRAS.497.5518A} {497, 5518}

\bibitem[\protect\citeauthoryear{{Astropy Collaboration} et~al.,}{{Astropy
  Collaboration} et~al.}{2013}]{astropy:2013}
{Astropy Collaboration} et~al., 2013, \mn@doi [\aap]
  {10.1051/0004-6361/201322068}, \href
  {http://adsabs.harvard.edu/abs/2013A%26A...558A..33A} {558, A33}

\bibitem[\protect\citeauthoryear{{Astropy Collaboration} et~al.,}{{Astropy
  Collaboration} et~al.}{2018}]{astropy:2018}
{Astropy Collaboration} et~al., 2018, \mn@doi [\aj] {10.3847/1538-3881/aabc4f},
  \href {https://ui.adsabs.harvard.edu/abs/2018AJ....156..123A} {156, 123}

\bibitem[\protect\citeauthoryear{{Bartos}, {Countryman}, {Finley}, {Blaufuss},
  {Corley}, {Marka}, {Marka}  \& {IceCube Collaboration}}{{Bartos}
  et~al.}{2017a}]{GCNicecube}
{Bartos} I.,  {Countryman} S.,  {Finley} C.,  {Blaufuss} E.,  {Corley} R.,
  {Marka} Z.,  {Marka} S.,   {IceCube Collaboration} 2017a, GRB Coordinates
  Network, \href {https://ui.adsabs.harvard.edu/abs/2017GCN.21694....1B}
  {21694}

\bibitem[\protect\citeauthoryear{{Bartos}, {Countryman}, {Finley}, {Blaufuss},
  {Corley}, {Marka}, {Marka}  \& {IceCube Collaboration}}{{Bartos}
  et~al.}{2017b}]{g299232Icecubeb}
{Bartos} I.,  {Countryman} S.,  {Finley} C.,  {Blaufuss} E.,  {Corley} R.,
  {Marka} Z.,  {Marka} S.,   {IceCube Collaboration} 2017b, GRB Coordinates
  Network, \href {https://ui.adsabs.harvard.edu/abs/2017GCN.21698....1B}
  {21698}

\bibitem[\protect\citeauthoryear{{Bell} et~al.,}{{Bell} et~al.}{2014}]{Bell14}
{Bell} M.~E.,  et~al., 2014, \mn@doi [\mnras] {10.1093/mnras/stt2200}, \href
  {https://ui.adsabs.harvard.edu/abs/2014MNRAS.438..352B} {438, 352}

\bibitem[\protect\citeauthoryear{{Boersma} et~al.,}{{Boersma}
  et~al.}{2021}]{Boersma2021}
{Boersma} O.,  et~al., 2021, arXiv e-prints, \href
  {https://ui.adsabs.harvard.edu/abs/2021arXiv210404280B} {p. arXiv:2104.04280}

\bibitem[\protect\citeauthoryear{{Broderick}, {Jonker}, {Fender}, {Rowlinson},
  {Wijers}  \& {Stappers}}{{Broderick} et~al.}{2015a}]{Broderick15a}
{Broderick} J.,  {Jonker} P.~G.,  {Fender} R.~P.,  {Rowlinson} A.,  {Wijers}
  R.~A.~M.~J.,   {Stappers} B.~W.,  2015a, GRB Coordinates Network, \href
  {https://ui.adsabs.harvard.edu/abs/2015GCN.18364....1B} {18364}

\bibitem[\protect\citeauthoryear{{Broderick}, {Jonker}, {Fender}, {Rowlinson},
  {Wijers}  \& {Stappers}}{{Broderick} et~al.}{2015b}]{Broderick15b}
{Broderick} J.,  {Jonker} P.~G.,  {Fender} R.~P.,  {Rowlinson} A.,  {Wijers}
  R.~A.~M.~J.,   {Stappers} B.~W.,  2015b, GRB Coordinates Network, \href
  {https://ui.adsabs.harvard.edu/abs/2015GCN.18424....1B} {18424}

\bibitem[\protect\citeauthoryear{{Broderick} et~al.,}{{Broderick}
  et~al.}{2017}]{Broderick17a}
{Broderick} J.~W.,  et~al., 2017, GRB Coordinates Network, \href
  {https://ui.adsabs.harvard.edu/abs/2017GCN.20953....1B} {20953}

\bibitem[\protect\citeauthoryear{{Broderick} et~al.,}{{Broderick}
  et~al.}{2020}]{Broderick20}
{Broderick} J.~W.,  et~al., 2020, \mn@doi [\mnras] {10.1093/mnras/staa950},
  \href {https://ui.adsabs.harvard.edu/abs/2020MNRAS.494.5110B} {494, 5110}

\bibitem[\protect\citeauthoryear{{Carbone} et~al.,}{{Carbone}
  et~al.}{2016}]{Carbone16}
{Carbone} D.,  et~al., 2016, \mn@doi [\mnras] {10.1093/mnras/stw539}, \href
  {https://ui.adsabs.harvard.edu/abs/2016MNRAS.459.3161C} {459, 3161}

\bibitem[\protect\citeauthoryear{{Carbone} et~al.,}{{Carbone}
  et~al.}{2018}]{pyse}
{Carbone} D.,  et~al., 2018, \mn@doi [Astronomy and Computing]
  {10.1016/j.ascom.2018.02.003}, \href
  {https://ui.adsabs.harvard.edu/abs/2018A&C....23...92C} {23, 92}

\bibitem[\protect\citeauthoryear{{Cendes} et~al.,}{{Cendes}
  et~al.}{2014}]{Cendes14}
{Cendes} Y.,  et~al., 2014, arXiv e-prints, \href
  {https://ui.adsabs.harvard.edu/abs/2014arXiv1412.3986C} {p. arXiv:1412.3986}

\bibitem[\protect\citeauthoryear{{Chu}, {Howell}, {Rowlinson}, {Gao}, {Zhang},
  {Tingay}, {Bo{\"e}r}  \& {Wen}}{{Chu} et~al.}{2016}]{Chu16}
{Chu} Q.,  {Howell} E.~J.,  {Rowlinson} A.,  {Gao} H.,  {Zhang} B.,  {Tingay}
  S.~J.,  {Bo{\"e}r} M.,   {Wen} L.,  2016, \mn@doi [\mnras]
  {10.1093/mnras/stw576}, \href
  {https://ui.adsabs.harvard.edu/abs/2016MNRAS.459..121C} {459, 121}

\bibitem[\protect\citeauthoryear{{Corsi} et~al.,}{{Corsi}
  et~al.}{2018}]{Corsi18}
{Corsi} A.,  et~al., 2018, \mn@doi [\apjl] {10.3847/2041-8213/aacdfd}, \href
  {https://ui.adsabs.harvard.edu/abs/2018ApJ...861L..10C} {861, L10}

\bibitem[\protect\citeauthoryear{{Coughlin} et~al.,}{{Coughlin}
  et~al.}{2019}]{Coughlin2019}
{Coughlin} M.~W.,  et~al., 2019, \mn@doi [\apjl] {10.3847/2041-8213/ab4ad8},
  \href {https://ui.adsabs.harvard.edu/abs/2019ApJ...885L..19C} {885, L19}

\bibitem[\protect\citeauthoryear{{Cowperthwaite} et~al.,}{{Cowperthwaite}
  et~al.}{2017}]{Cowperthwaite17}
{Cowperthwaite} P.~S.,  et~al., 2017, \mn@doi [\apjl]
  {10.3847/2041-8213/aa8fc7}, \href
  {https://ui.adsabs.harvard.edu/abs/2017ApJ...848L..17C} {848, L17}

\bibitem[\protect\citeauthoryear{{D'Orazio}, {Levin}, {Murray}  \&
  {Price}}{{D'Orazio} et~al.}{2016}]{Dorazio16}
{D'Orazio} D.~J.,  {Levin} J.,  {Murray} N.~W.,   {Price} L.,  2016, \mn@doi
  [\prd] {10.1103/PhysRevD.94.023001}, \href
  {https://ui.adsabs.harvard.edu/abs/2016PhRvD..94b3001D} {94, 023001}

\bibitem[\protect\citeauthoryear{{Dobie} et~al.,}{{Dobie}
  et~al.}{2018}]{Dobie18}
{Dobie} D.,  et~al., 2018, \mn@doi [\apjl] {10.3847/2041-8213/aac105}, \href
  {https://ui.adsabs.harvard.edu/abs/2018ApJ...858L..15D} {858, L15}

\bibitem[\protect\citeauthoryear{{Dobie} et~al.,}{{Dobie}
  et~al.}{2019}]{Dobie19}
{Dobie} D.,  et~al., 2019, \mn@doi [\apjl] {10.3847/2041-8213/ab59db}, \href
  {https://ui.adsabs.harvard.edu/abs/2019ApJ...887L..13D} {887, L13}

\bibitem[\protect\citeauthoryear{{Emery} et~al.,}{{Emery}
  et~al.}{2017}]{EmeryGCN}
{Emery} S.~W.~K.,  et~al., 2017, GRB Coordinates Network, \href
  {https://ui.adsabs.harvard.edu/abs/2017GCN.21733....1E} {21733}

\bibitem[\protect\citeauthoryear{{Falcke} \& {Rezzolla}}{{Falcke} \&
  {Rezzolla}}{2014}]{FalckeRezzolla}
{Falcke} H.,  {Rezzolla} L.,  2014, \mn@doi [\aap]
  {10.1051/0004-6361/201321996}, \href
  {https://ui.adsabs.harvard.edu/abs/2014A&A...562A.137F} {562, A137}

\bibitem[\protect\citeauthoryear{{Feng} et~al.,}{{Feng} et~al.}{2017}]{Feng17}
{Feng} L.,  et~al., 2017, \mn@doi [\aj] {10.3847/1538-3881/153/3/98}, \href
  {https://ui.adsabs.harvard.edu/abs/2017AJ....153...98F} {153, 98}

\bibitem[\protect\citeauthoryear{{Ghirlanda} et~al.,}{{Ghirlanda}
  et~al.}{2019}]{Ghirlanda19}
{Ghirlanda} G.,  et~al., 2019, \mn@doi [Science] {10.1126/science.aau8815},
  \href {https://ui.adsabs.harvard.edu/abs/2019Sci...363..968G} {363, 968}

\bibitem[\protect\citeauthoryear{{Gourdji}, {Rowlinson}, {Wijers}  \&
  {Goldstein}}{{Gourdji} et~al.}{2020}]{Gourdji2020}
{Gourdji} K.,  {Rowlinson} A.,  {Wijers} R.~A.~M.~J.,   {Goldstein} A.,  2020,
  \mn@doi [\mnras] {10.1093/mnras/staa2128}, \href
  {https://ui.adsabs.harvard.edu/abs/2020MNRAS.497.3131G} {497, 3131}

\bibitem[\protect\citeauthoryear{{Hajela} et~al.,}{{Hajela}
  et~al.}{2019}]{Hajela19}
{Hajela} A.,  et~al., 2019, \mn@doi [\apjl] {10.3847/2041-8213/ab5226}, \href
  {https://ui.adsabs.harvard.edu/abs/2019ApJ...886L..17H} {886, L17}

\bibitem[\protect\citeauthoryear{{Hallinan} et~al.,}{{Hallinan}
  et~al.}{2017}]{Hallinan17}
{Hallinan} G.,  et~al., 2017, \mn@doi [Science] {10.1126/science.aap9855},
  \href {https://ui.adsabs.harvard.edu/abs/2017Sci...358.1579H} {358, 1579}

\bibitem[\protect\citeauthoryear{{Hansen} \& {Lyutikov}}{{Hansen} \&
  {Lyutikov}}{2001}]{Hansen2001}
{Hansen} B. M.~S.,  {Lyutikov} M.,  2001, \mn@doi [\mnras]
  {10.1046/j.1365-8711.2001.04103.x}, \href
  {https://ui.adsabs.harvard.edu/abs/2001MNRAS.322..695H} {322, 695}

\bibitem[\protect\citeauthoryear{{Hotokezaka}, {Nissanke}, {Hallinan}, {Lazio},
  {Nakar}  \& {Piran}}{{Hotokezaka} et~al.}{2016}]{Hotokezaka16}
{Hotokezaka} K.,  {Nissanke} S.,  {Hallinan} G.,  {Lazio} T.~J.~W.,  {Nakar}
  E.,   {Piran} T.,  2016, \mn@doi [\apj] {10.3847/0004-637X/831/2/190}, \href
  {https://ui.adsabs.harvard.edu/abs/2016ApJ...831..190H} {831, 190}

\bibitem[\protect\citeauthoryear{{Hyman}, {Wijnands}, {Lazio}, {Pal},
  {Starling}, {Kassim}  \& {Ray}}{{Hyman} et~al.}{2009}]{Hyman2009}
{Hyman} S.~D.,  {Wijnands} R.,  {Lazio} T. J.~W.,  {Pal} S.,  {Starling} R.,
  {Kassim} N.~E.,   {Ray} P.~S.,  2009, \mn@doi [\apj]
  {10.1088/0004-637X/696/1/280}, \href
  {https://ui.adsabs.harvard.edu/abs/2009ApJ...696..280H} {696, 280}

\bibitem[\protect\citeauthoryear{{Intema}, {Jagannathan}, {Mooley}  \&
  {Frail}}{{Intema} et~al.}{2017}]{Intema17}
{Intema} H.~T.,  {Jagannathan} P.,  {Mooley} K.~P.,   {Frail} D.~A.,  2017,
  \mn@doi [\aap] {10.1051/0004-6361/201628536}, \href
  {https://ui.adsabs.harvard.edu/abs/2017A&A...598A..78I} {598, A78}

\bibitem[\protect\citeauthoryear{{Jaeger}, {Hyman}, {Kassim}  \&
  {Lazio}}{{Jaeger} et~al.}{2012}]{Jaeger12}
{Jaeger} T.~R.,  {Hyman} S.~D.,  {Kassim} N.~E.,   {Lazio} T.~J.~W.,  2012,
  \mn@doi [\aj] {10.1088/0004-6256/143/4/96}, \href
  {https://ui.adsabs.harvard.edu/abs/2012AJ....143...96J} {143, 96}

\bibitem[\protect\citeauthoryear{{Jonker} et~al.,}{{Jonker}
  et~al.}{2017}]{JonkerGCN}
{Jonker} P.~G.,  et~al., 2017, GRB Coordinates Network, \href
  {https://ui.adsabs.harvard.edu/abs/2017GCN.21771....1S} {21771}

\bibitem[\protect\citeauthoryear{{LIGO Scientific Collaboration/Virgo
  Collaboration}}{{LIGO Scientific Collaboration/Virgo
  Collaboration}}{2017}]{g299232gcn}
{LIGO Scientific Collaboration/Virgo Collaboration} 2017, GRB Coordinates
  Network, 21693

\bibitem[\protect\citeauthoryear{{Lazio} et~al.,}{{Lazio}
  et~al.}{2010}]{Lazio10}
{Lazio} T. J.~W.,  et~al., 2010, \mn@doi [\aj] {10.1088/0004-6256/140/6/1995},
  \href {https://ui.adsabs.harvard.edu/abs/2010AJ....140.1995L} {140, 1995}

\bibitem[\protect\citeauthoryear{{Margalit} \& {Metzger}}{{Margalit} \&
  {Metzger}}{2017}]{MargalitMetzger17}
{Margalit} B.,  {Metzger} B.~D.,  2017, \mn@doi [\apjl]
  {10.3847/2041-8213/aa991c}, \href
  {https://ui.adsabs.harvard.edu/abs/2017ApJ...850L..19M} {850, L19}

\bibitem[\protect\citeauthoryear{{Margutti} et~al.,}{{Margutti}
  et~al.}{2018}]{Margutti18}
{Margutti} R.,  et~al., 2018, \mn@doi [\apjl] {10.3847/2041-8213/aab2ad}, \href
  {https://ui.adsabs.harvard.edu/abs/2018ApJ...856L..18M} {856, L18}

\bibitem[\protect\citeauthoryear{{M{\'e}sz{\'a}ros} \&
  {Rees}}{{M{\'e}sz{\'a}ros} \& {Rees}}{1997}]{MeszarosRees97}
{M{\'e}sz{\'a}ros} P.,  {Rees} M.~J.,  1997, \mn@doi [\apj] {10.1086/303625},
  \href {https://ui.adsabs.harvard.edu/abs/1997ApJ...476..232M} {476, 232}

\bibitem[\protect\citeauthoryear{{Minns} \& {Riley}}{{Minns} \&
  {Riley}}{2000}]{MinnsRiley2000}
{Minns} A.~R.,  {Riley} J.~M.,  2000, \mn@doi [\mnras]
  {10.1046/j.1365-8711.2000.03506.x}, \href
  {https://ui.adsabs.harvard.edu/abs/2000MNRAS.315..839M} {315, 839}

\bibitem[\protect\citeauthoryear{{Mohan} \& {Rafferty}}{{Mohan} \&
  {Rafferty}}{2015}]{pybdsf}
{Mohan} N.,  {Rafferty} D.,  2015, {PyBDSF: Python Blob Detection and Source
  Finder} (\mn@eprint {ascl} {1502.007})

\bibitem[\protect\citeauthoryear{{Mooley} et~al.,}{{Mooley}
  et~al.}{2016}]{Mooley16}
{Mooley} K.~P.,  et~al., 2016, \mn@doi [\apj] {10.3847/0004-637X/818/2/105},
  \href {https://ui.adsabs.harvard.edu/abs/2016ApJ...818..105M} {818, 105}

\bibitem[\protect\citeauthoryear{{Mooley} et~al.,}{{Mooley}
  et~al.}{2018a}]{Mooley18a}
{Mooley} K.~P.,  et~al., 2018a, \mn@doi [\nat] {10.1038/nature25452}, \href
  {https://ui.adsabs.harvard.edu/abs/2018Natur.554..207M} {554, 207}

\bibitem[\protect\citeauthoryear{{Mooley} et~al.,}{{Mooley}
  et~al.}{2018b}]{Mooley18b}
{Mooley} K.~P.,  et~al., 2018b, \mn@doi [\nat] {10.1038/s41586-018-0486-3},
  \href {https://ui.adsabs.harvard.edu/abs/2018Natur.561..355M} {561, 355}

\bibitem[\protect\citeauthoryear{{Mooley} et~al.,}{{Mooley}
  et~al.}{2018c}]{Mooley18c}
{Mooley} K.~P.,  et~al., 2018c, \mn@doi [\apjl] {10.3847/2041-8213/aaeda7},
  \href {https://ui.adsabs.harvard.edu/abs/2018ApJ...868L..11M} {868, L11}

\bibitem[\protect\citeauthoryear{{Murphy} et~al.,}{{Murphy}
  et~al.}{2017}]{Murphy2017}
{Murphy} T.,  et~al., 2017, \mn@doi [\mnras] {10.1093/mnras/stw3087}, \href
  {https://ui.adsabs.harvard.edu/abs/2017MNRAS.466.1944M} {466, 1944}

\bibitem[\protect\citeauthoryear{{Nakar} \& {Piran}}{{Nakar} \&
  {Piran}}{2011}]{NakarPiran11}
{Nakar} E.,  {Piran} T.,  2011, \mn@doi [\nat] {10.1038/nature10365}, \href
  {https://ui.adsabs.harvard.edu/abs/2011Natur.478...82N} {478, 82}

\bibitem[\protect\citeauthoryear{{Offringa}, {van de Gronde}  \&
  {Roerdink}}{{Offringa} et~al.}{2012}]{Offringa12}
{Offringa} A.~R.,  {van de Gronde} J.~J.,   {Roerdink} J.~B.~T.~M.,  2012,
  \mn@doi [\aap] {10.1051/0004-6361/201118497}, \href
  {https://ui.adsabs.harvard.edu/abs/2012A&A...539A..95O} {539, A95}

\bibitem[\protect\citeauthoryear{{Offringa} et~al.,}{{Offringa}
  et~al.}{2014}]{wsclean14}
{Offringa} A.~R.,  et~al., 2014, \mn@doi [\mnras] {10.1093/mnras/stu1368},
  \href {https://ui.adsabs.harvard.edu/abs/2014MNRAS.444..606O} {444, 606}

\bibitem[\protect\citeauthoryear{{Paczynski} \& {Rhoads}}{{Paczynski} \&
  {Rhoads}}{1993}]{PaczynskiRhoads93}
{Paczynski} B.,  {Rhoads} J.~E.,  1993, \mn@doi [\apjl] {10.1086/187102}, \href
  {https://ui.adsabs.harvard.edu/abs/1993ApJ...418L...5P} {418, L5}

\bibitem[\protect\citeauthoryear{{Page} et~al.,}{{Page}
  et~al.}{2020}]{Page2020}
{Page} K.~L.,  et~al., 2020, \mn@doi [\mnras] {10.1093/mnras/staa3032}, \href
  {https://ui.adsabs.harvard.edu/abs/2020MNRAS.499.3459P} {499, 3459}

\bibitem[\protect\citeauthoryear{{Pandey} \& {Lofar Eor Group}}{{Pandey} \&
  {Lofar Eor Group}}{2014}]{Pandey3C196}
{Pandey} V.,  {Lofar Eor Group} 2014, in Exascale Radio Astronomy. p. 10401

\bibitem[\protect\citeauthoryear{{Pastor-Marazuela} et~al.,}{{Pastor-Marazuela}
  et~al.}{2020}]{pastor21}
{Pastor-Marazuela} I.,  et~al., 2020, arXiv e-prints, \href
  {https://ui.adsabs.harvard.edu/abs/2020arXiv201208348P} {p. arXiv:2012.08348}

\bibitem[\protect\citeauthoryear{{Piro}}{{Piro}}{2012}]{piro2012}
{Piro} A.~L.,  2012, \mn@doi [\apj] {10.1088/0004-637X/755/1/80}, \href
  {https://ui.adsabs.harvard.edu/abs/2012ApJ...755...80P} {755, 80}

\bibitem[\protect\citeauthoryear{{Pleunis} et~al.,}{{Pleunis}
  et~al.}{2021}]{pleunis21}
{Pleunis} Z.,  et~al., 2021, \mn@doi [\apjl] {10.3847/2041-8213/abec72}, \href
  {https://ui.adsabs.harvard.edu/abs/2021ApJ...911L...3P} {911, L3}

\bibitem[\protect\citeauthoryear{{Polisensky} et~al.,}{{Polisensky}
  et~al.}{2016}]{Polisensky16}
{Polisensky} E.,  et~al., 2016, \mn@doi [\apj] {10.3847/0004-637X/832/1/60},
  \href {https://ui.adsabs.harvard.edu/abs/2016ApJ...832...60P} {832, 60}

\bibitem[\protect\citeauthoryear{{Pshirkov} \& {Postnov}}{{Pshirkov} \&
  {Postnov}}{2010}]{PshirkovPostnov}
{Pshirkov} M.~S.,  {Postnov} K.~A.,  2010, \mn@doi [\apss]
  {10.1007/s10509-010-0395-x}, \href
  {https://ui.adsabs.harvard.edu/abs/2010Ap&SS.330...13P} {330, 13}

\bibitem[\protect\citeauthoryear{{Resmi} et~al.,}{{Resmi}
  et~al.}{2018}]{Resmi18}
{Resmi} L.,  et~al., 2018, \mn@doi [\apj] {10.3847/1538-4357/aae1a6}, \href
  {https://ui.adsabs.harvard.edu/abs/2018ApJ...867...57R} {867, 57}

\bibitem[\protect\citeauthoryear{{Rowlinson} \& {Anderson}}{{Rowlinson} \&
  {Anderson}}{2019}]{RowlinsonAnderson20}
{Rowlinson} A.,  {Anderson} G.~E.,  2019, \mn@doi [\mnras]
  {10.1093/mnras/stz2295}, \href
  {https://ui.adsabs.harvard.edu/abs/2019MNRAS.489.3316R} {489, 3316}

\bibitem[\protect\citeauthoryear{{Rowlinson} et~al.,}{{Rowlinson}
  et~al.}{2016}]{Rowlinson16}
{Rowlinson} A.,  et~al., 2016, \mn@doi [\mnras] {10.1093/mnras/stw451}, \href
  {https://ui.adsabs.harvard.edu/abs/2016MNRAS.458.3506R} {458, 3506}

\bibitem[\protect\citeauthoryear{{Rowlinson} et~al.,}{{Rowlinson}
  et~al.}{2017a}]{Rowlinson17a}
{Rowlinson} A.,  et~al., 2017a, GRB Coordinates Network, \href
  {https://ui.adsabs.harvard.edu/abs/2017GCN.20372....1R} {20372}

\bibitem[\protect\citeauthoryear{{Rowlinson} et~al.,}{{Rowlinson}
  et~al.}{2017b}]{Rowlinson17b}
{Rowlinson} A.,  et~al., 2017b, GRB Coordinates Network, \href
  {https://ui.adsabs.harvard.edu/abs/2017GCN.21355....1R} {21355}

\bibitem[\protect\citeauthoryear{{Rowlinson} et~al.,}{{Rowlinson}
  et~al.}{2019}]{Rowlinson19}
{Rowlinson} A.,  et~al., 2019, \mn@doi [\mnras] {10.1093/mnras/stz2866}, \href
  {https://ui.adsabs.harvard.edu/abs/2019MNRAS.490.3483R} {490, 3483}

\bibitem[\protect\citeauthoryear{{Rowlinson} et~al.,}{{Rowlinson}
  et~al.}{2020}]{Rowlinson20}
{Rowlinson} A.,  et~al., 2020, arXiv e-prints, \href
  {https://ui.adsabs.harvard.edu/abs/2020arXiv200812657R} {p. arXiv:2008.12657}

\bibitem[\protect\citeauthoryear{{Scaife} \& {Heald}}{{Scaife} \&
  {Heald}}{2012}]{ScaifeHeald12}
{Scaife} A. M.~M.,  {Heald} G.~H.,  2012, \mn@doi [\mnras]
  {10.1111/j.1745-3933.2012.01251.x}, \href
  {https://ui.adsabs.harvard.edu/abs/2012MNRAS.423L..30S} {423, L30}

\bibitem[\protect\citeauthoryear{{Scheers}}{{Scheers}}{2011}]{scheers11}
{Scheers} L.~H.~A.,  2011, PhD thesis, University of Amsterdam

\bibitem[\protect\citeauthoryear{{Shimwell} et~al.,}{{Shimwell}
  et~al.}{2019}]{Shimwell19}
{Shimwell} T.~W.,  et~al., 2019, \mn@doi [\aap] {10.1051/0004-6361/201833559},
  \href {https://ui.adsabs.harvard.edu/abs/2019A&A...622A...1S} {622, A1}

\bibitem[\protect\citeauthoryear{{Smirnov} \& {Tasse}}{{Smirnov} \&
  {Tasse}}{2015}]{SmirnovTasse15}
{Smirnov} O.~M.,  {Tasse} C.,  2015, \mn@doi [\mnras] {10.1093/mnras/stv418},
  \href {https://ui.adsabs.harvard.edu/abs/2015MNRAS.449.2668S} {449, 2668}

\bibitem[\protect\citeauthoryear{{Stewart} et~al.,}{{Stewart}
  et~al.}{2016}]{Stewart16}
{Stewart} A.~J.,  et~al., 2016, \mn@doi [\mnras] {10.1093/mnras/stv2797}, \href
  {https://ui.adsabs.harvard.edu/abs/2016MNRAS.456.2321S} {456, 2321}

\bibitem[\protect\citeauthoryear{Swinbank et~al.,}{Swinbank
  et~al.}{2015}]{Swinbank2015}
Swinbank J.~D.,  et~al., 2015, \mn@doi [Astronomy and Computing]
  {https://doi.org/10.1016/j.ascom.2015.03.002}, 11, 25

\bibitem[\protect\citeauthoryear{{Tasse}}{{Tasse}}{2014a}]{Tasse14a}
{Tasse} C.,  2014a, arXiv e-prints, \href
  {https://ui.adsabs.harvard.edu/abs/2014arXiv1410.8706T} {p. arXiv:1410.8706}

\bibitem[\protect\citeauthoryear{{Tasse}}{{Tasse}}{2014b}]{Tasse14b}
{Tasse} C.,  2014b, \mn@doi [\aap] {10.1051/0004-6361/201423503}, \href
  {https://ui.adsabs.harvard.edu/abs/2014A&A...566A.127T} {566, A127}

\bibitem[\protect\citeauthoryear{{Tasse} et~al.,}{{Tasse}
  et~al.}{2018}]{Tasse18}
{Tasse} C.,  et~al., 2018, \mn@doi [\aap] {10.1051/0004-6361/201731474}, \href
  {https://ui.adsabs.harvard.edu/abs/2018A&A...611A..87T} {611, A87}

\bibitem[\protect\citeauthoryear{{Totani}}{{Totani}}{2013}]{totani13}
{Totani} T.,  2013, \mn@doi [\pasj] {10.1093/pasj/65.5.L12}, \href
  {https://ui.adsabs.harvard.edu/abs/2013PASJ...65L..12T} {65, L12}

\bibitem[\protect\citeauthoryear{{Troja} et~al.,}{{Troja}
  et~al.}{2018}]{Troja18}
{Troja} E.,  et~al., 2018, \mn@doi [\mnras] {10.1093/mnrasl/sly061}, \href
  {https://ui.adsabs.harvard.edu/abs/2018MNRAS.478L..18T} {478, L18}

\bibitem[\protect\citeauthoryear{{Usov} \& {Katz}}{{Usov} \&
  {Katz}}{2000}]{UsovKatz}
{Usov} V.~V.,  {Katz} J.~I.,  2000, \aap, \href
  {https://ui.adsabs.harvard.edu/abs/2000A&A...364..655U} {364, 655}

\bibitem[\protect\citeauthoryear{{de Gasperin} et~al.,}{{de Gasperin}
  et~al.}{2019}]{prefactor}
{de Gasperin} F.,  et~al., 2019, \mn@doi [\aap] {10.1051/0004-6361/201833867},
  \href {https://ui.adsabs.harvard.edu/abs/2019A&A...622A...5D} {622, A5}

\bibitem[\protect\citeauthoryear{{de Ruiter}, {Willis}  \& {Arp}}{{de Ruiter}
  et~al.}{1977}]{deRuiter77}
{de Ruiter} H.~R.,  {Willis} A.~G.,   {Arp} H.~C.,  1977, \aaps, \href
  {https://ui.adsabs.harvard.edu/abs/1977A&AS...28..211D} {28, 211}

\bibitem[\protect\citeauthoryear{{de Ruiter}, {Leseigneur}, {Rowlinson},
  {Wijers}, {Drabent}, {Intema}  \& {Shimwell}}{{de Ruiter}
  et~al.}{2021}]{deRuiter21}
{de Ruiter} I.,  {Leseigneur} G.,  {Rowlinson} A.,  {Wijers} R. A.~M.~J.,
  {Drabent} A.,  {Intema} H.~T.,   {Shimwell} T.~W.,  2021, \mn@doi [\mnras]
  {10.1093/mnras/stab2695}, \href
  {https://ui.adsabs.harvard.edu/abs/2021MNRAS.508.2412D} {508, 2412}

\bibitem[\protect\citeauthoryear{{van Haarlem} et~al.,}{{van Haarlem}
  et~al.}{2013}]{LOFAR13}
{van Haarlem} M.~P.,  et~al., 2013, \mn@doi [\aap]
  {10.1051/0004-6361/201220873}, \href
  {https://ui.adsabs.harvard.edu/abs/2013A&A...556A...2V} {556, A2}

\makeatother
\end{thebibliography}


\captionsetup[subfigure]{labelformat=empty}
\captionsetup{labelfont=bf}
\appendix
\onecolumn
\section{Transient candidate example classifications}
\noindent%
\begin{minipage}{\linewidth}
    \captionof{figure}{Examples for each of the five new-source candidate types listed and described in section \ref{meth:blind}. The labels on the $x$ and $y$ axes are in pixel units. All plots are centered on the new-source candidate. The red circle denotes the source position error radius and the yellow ellipse corresponds to the 2D Gaussian source fit as determined by \textsc{TraP}. The restoring beam of each image is shown in the bottom left corner and the pixel size is 5\arcsec. }
    \begin{subfigure}[t]{0.31\textwidth}
        \centering
        \includegraphics[width=1.07\textwidth]{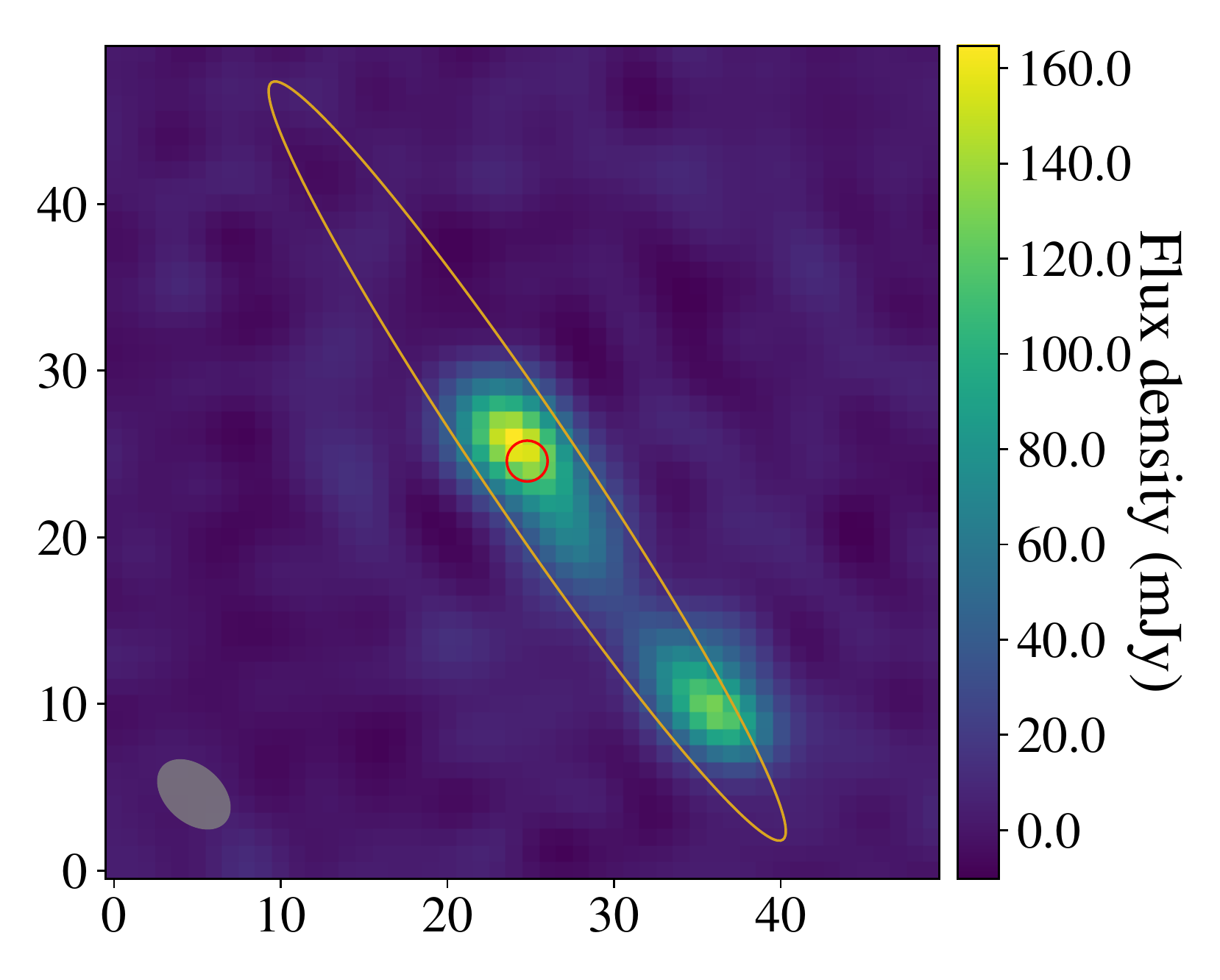}
        \caption{Type (i) extended source with differing fit parameters between epochs, which prevent association. }
    \end{subfigure}
    \hfill
    \begin{subfigure}[t]{0.33\textwidth}
        \centering
    	\includegraphics[width=\textwidth]{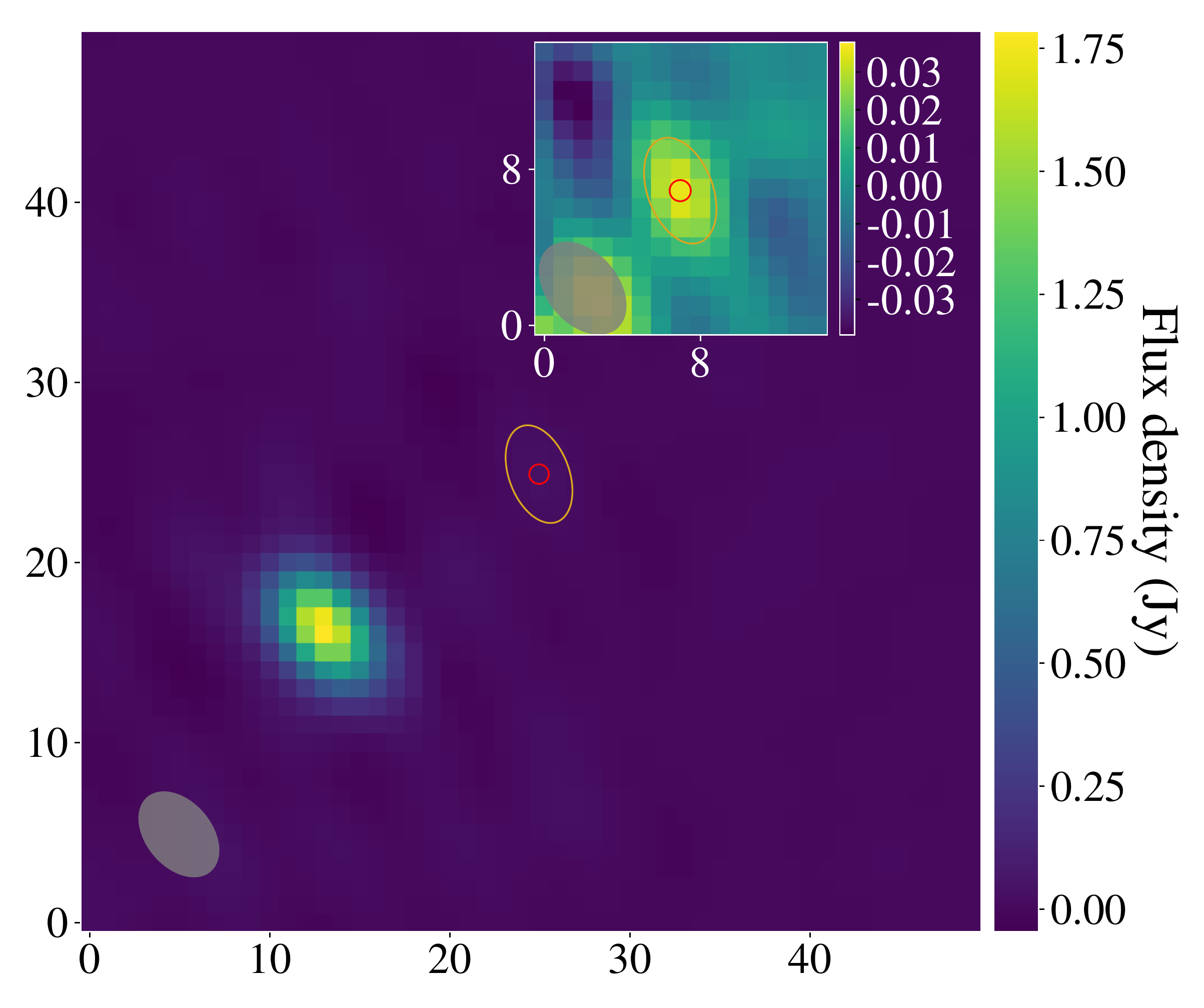}
        \caption{Type (ii) false positive caused by a sidelobe around a bright nearby source. The inset shows a zoom-in centered on the candidate.}
    \end{subfigure}
        \hfill
    \begin{subfigure}[t]{0.33\textwidth}
        \centering
    	\includegraphics[width=\textwidth]{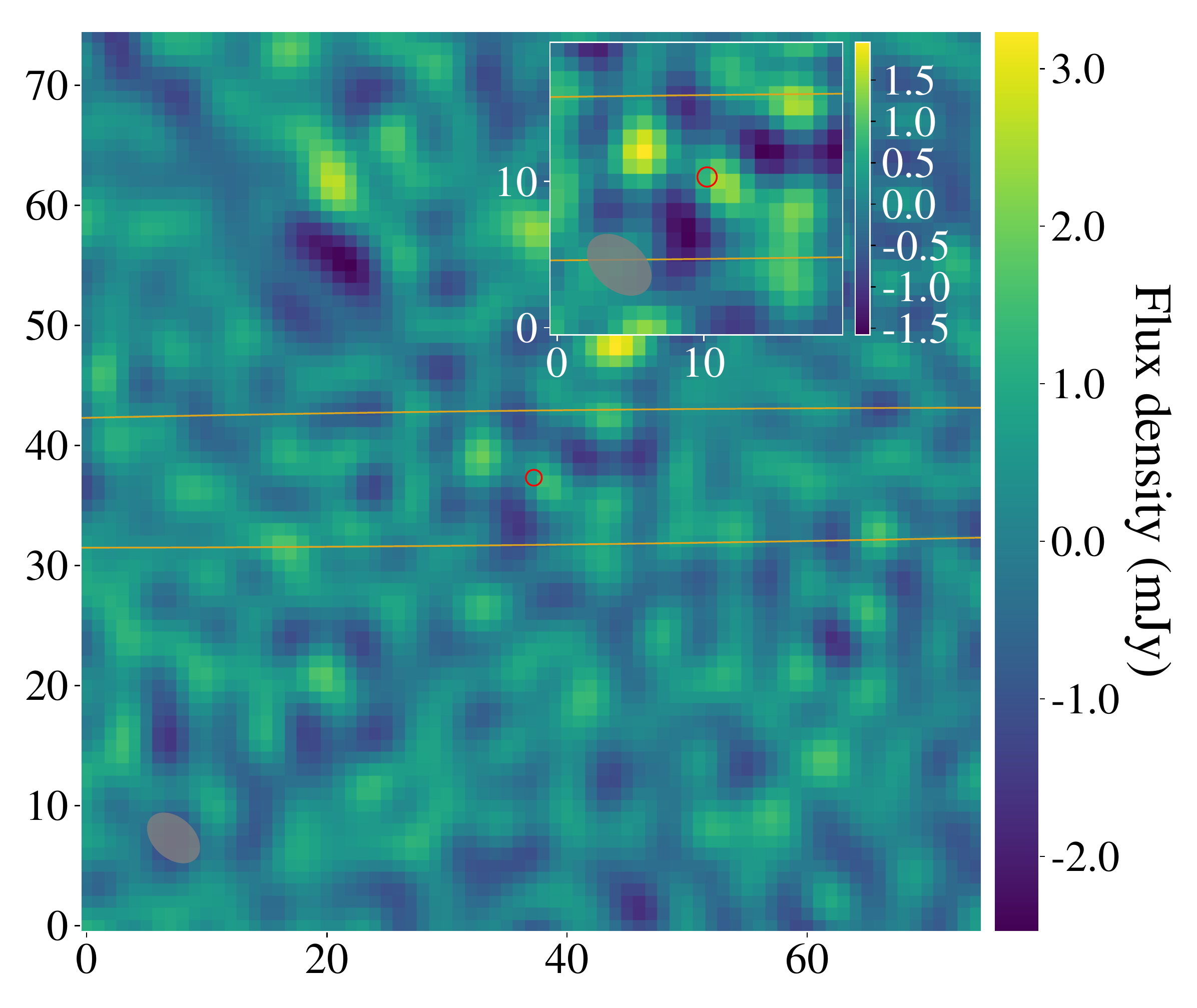}
        \caption{Type (iii) false positive caused by correlated noise. The major axis of the \textsc{TraP} fit is approximately 300 pixels long and is cut off here. The inset shows a zoom-in centered on the candidate.}
    \end{subfigure}
    \bigskip
        \begin{subfigure}[t]{\textwidth}
        \centering
    	\includegraphics[width=\textwidth]{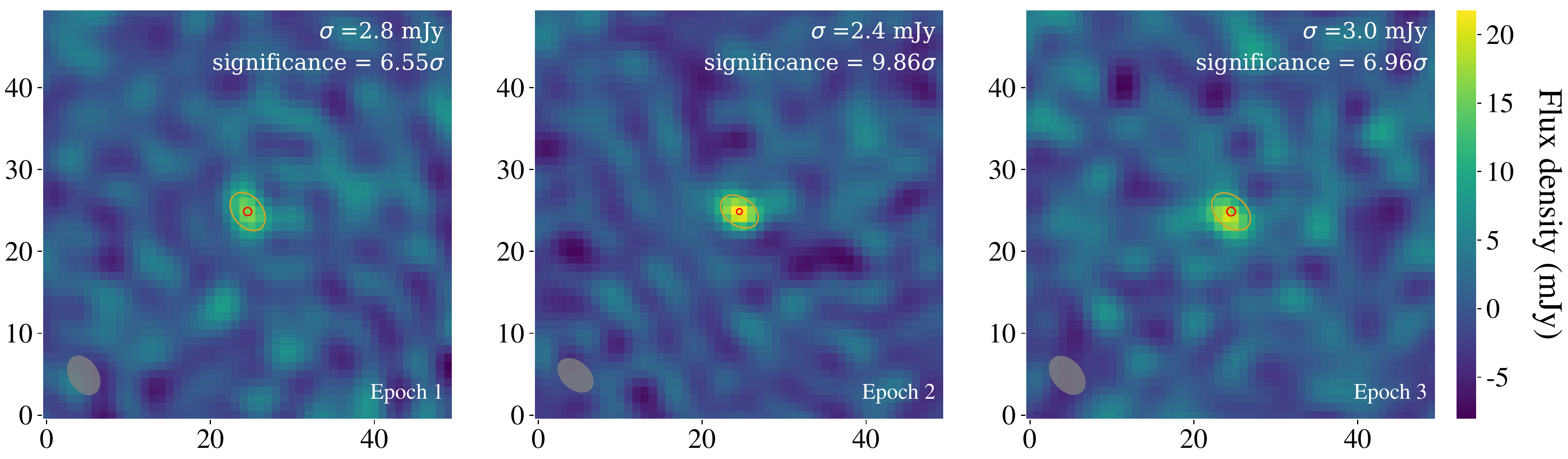}
        \caption{Type (iv): false positive caused by real sources with fluxes around the detection threshold and differing image sensitivity between epochs. In this example, the source was detected in the second epoch image (center), but due to poorer sensitivity in the first and third epoch images, was undetected in the other epochs. }
    \end{subfigure}
    \bigskip
        \begin{subfigure}[t]{\textwidth}
        \centering
    	\includegraphics[width=\textwidth]{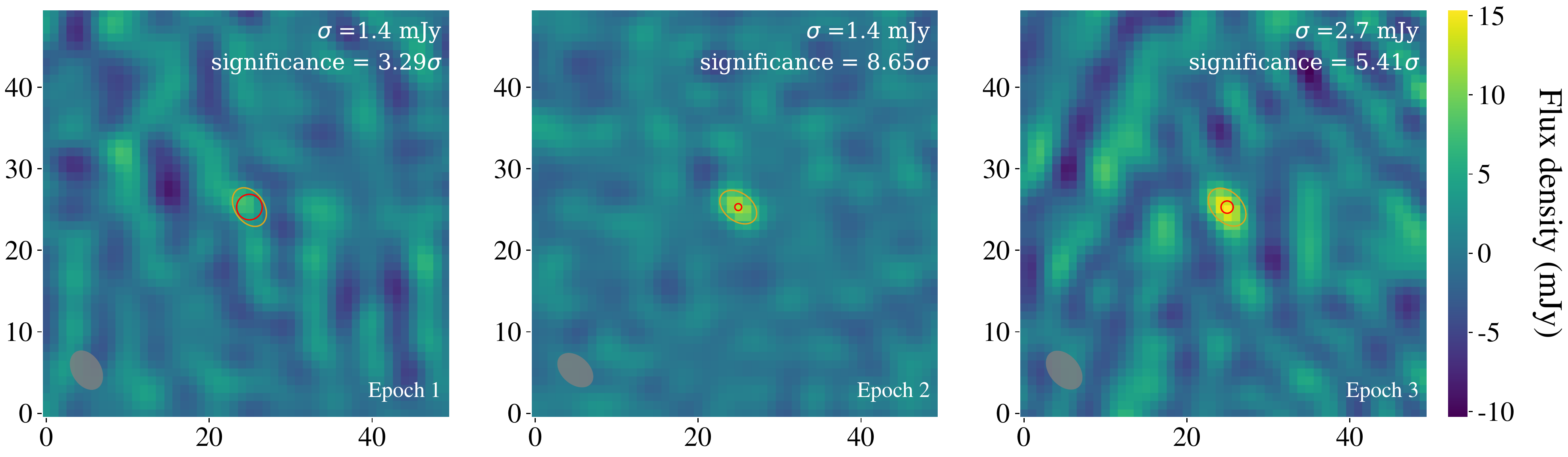}
        \caption{Type (v): Transient candidate that passed through all filters as well as visual inspection. The source is detected in the second epoch image (centre) and is still visible in the third epoch (right). The candidate was ruled out after finding that the normalized differences between the candidate's peak flux and the forced flux measurements of the other two images were insignificant (see Figure \ref{fig:flux_diff}).}
    \end{subfigure}
\label{fig:cand types}
\end{minipage}
\newpage
\section{Observation and source count tables}
    \centering
    \footnotesize
    \captionof{table}{Table of observations. The observation start times are shown and the mean image RMS is included.\newline
    $^a$Data calibrated using 3C48 instead of 3C196.\\
    $^b$Observations where one of the nine 25-minute scans is missing.
    \label{tab:obs}}
    \begin{filecontents*}{obs_table_g299232.csv}
RA (deg),DEC (deg),\thead{Epoch 1\\(UTC)},\thead{$\langle$RMS$_1\rangle$\\mJy\\beam$^{-1}$},\thead{Epoch 2\\(UTC)},\thead{$\langle$RMS$_2\rangle$\\mJy\\beam$^{-1}$},\thead{Epoch 3\\(UTC)},\thead{$\langle$RMS$_3\rangle$\\mJy\\beam$^{-1}$}
25.035625,34.56766667,$^a$2017-08-31 23:00:05,6.2,2017-09-24 21:00:05,5.4,2017-11-20 18:11:05,5.1
27.1051976779,37.9065470782,$^a$2017-08-31 23:00:05,2.6,2017-09-24 21:00:05,3.4,2017-11-20 18:11:05,2.7
28.1993407705,40.5175129993,$^a$2017-08-31 23:26:05,2.5,2017-09-24 21:26:05,3.0,2017-11-20 18:37:05,2.3
28.2,44.8,2017-08-30 23:00:05,2.8,2017-09-23 21:00:05,2.0,2017-11-18 18:11:05,2.7
29.3066435278,35.7780681577,$^a$2017-08-31 23:00:05,3.0,2017-09-24 21:00:05,2.8,2017-11-20 18:11:05,2.5
30.4780083982,38.3890340788,$^a$2017-08-31 23:00:05,2.3,2017-09-24 21:00:05,2.7,2017-11-20 18:11:05,2.2
31.320689341,78.2892389693,2017-09-03 01:26:05,2.1,2017-09-26 21:37:05,1.8,2017-11-24 18:37:05,3.5
31.3221270833,48.3484145366,2017-08-30 23:00:05,1.5,2017-09-23 21:00:05,1.7,2017-11-18 18:11:05,2.2
31.5020712805,50.8896844781,2017-08-30 23:26:05,1.6,2017-09-23 21:26:05,1.6,2017-11-18 18:37:05,2.0
31.5080893777,33.6495892372,$^a$2017-08-31 23:00:05,3.0,2017-09-24 21:00:05,3.6,2017-11-20 18:11:05,3.1
31.825,41.0,$^a$2017-08-31 23:26:05,2.4,2017-09-24 21:26:05,3.5,2017-11-20 18:37:05,2.2
31.990305671,81.0651263206,2017-09-03 01:26:05,2.3,2017-09-26 21:37:05,1.7,2017-11-24 18:37:05,3.0
32.4729327845,53.6023540758,2017-08-30 23:26:05,1.5,2017-09-23 21:26:05,1.9,$^b$2017-11-18 18:37:05,2.3
32.5183026084,56.2475307662,2017-09-04 00:00:05,2.6,2017-09-27 22:00:05,2.6,2017-11-25 18:11:05,2.8
32.6794542481,36.2605551584,$^a$2017-08-31 23:00:05,2.9,2017-09-24 21:00:05,3.5,2017-11-20 18:11:05,2.9
33.0841782361,43.6109659212,$^a$2017-08-31 23:26:05,2.9,2017-09-24 21:26:05,3.8,2017-11-20 18:37:05,2.6
33.4089275424,58.9838505214,2017-09-04 00:00:05,2.7,2017-09-27 22:00:05,2.7,2017-11-25 18:11:05,2.9
33.4320963565,61.7201702765,2017-09-04 00:26:05,2.7,2017-09-27 22:26:05,2.9,2017-11-25 18:37:05,3.4
34.0391514504,46.2998659386,2017-08-30 23:00:05,1.4,2017-09-23 21:00:05,1.6,$^b$2017-11-18 18:11:05,2.3
34.1914809934,38.8715210795,$^a$2017-08-31 23:26:05,3.0,2017-09-24 21:26:05,4.3,2017-11-20 18:37:05,3.1
34.4830207023,64.4564900316,2017-09-04 00:26:05,2.3,2017-09-27 22:26:05,2.8,2017-11-25 18:37:05,3.2
34.6356226701,67.260302694,2017-09-03 01:00:05,1.8,2017-09-26 21:11:05,1.7,2017-11-24 18:11:05,2.4
35.2485014015,48.9497989079,2017-08-30 23:00:05,1.5,2017-09-23 21:00:05,1.7,$^b$2017-11-18 18:11:05,2.1
35.4506592295,41.4824870007,$^a$2017-08-31 23:26:05,3.9,2017-09-24 21:26:05,5.7,2017-11-20 18:37:05,3.4
35.7800639102,51.725205134,2017-08-30 23:26:05,1.5,2017-09-23 21:26:05,1.7,2017-11-18 18:37:05,2.1
36.247673704,69.9729722917,2017-09-03 01:00:05,1.6,2017-09-26 21:11:05,1.5,2017-11-24 18:11:05,2.3
36.7509254142,54.4378747316,2017-08-30 23:26:05,1.9,2017-09-23 21:26:05,2.0,2017-11-18 18:37:05,2.2
36.7561758175,44.2513173406,2017-08-30 23:00:05,1.8,2017-09-23 21:00:05,2.0,2017-11-18 18:11:05,2.7
37.3378963963,57.1978446442,2017-09-04 00:00:05,2.9,2017-09-27 22:00:05,3.1,2017-11-25 18:11:05,3.4
37.9655257686,46.9012503099,2017-08-30 23:00:05,1.6,2017-09-23 21:00:05,1.7,2017-11-18 18:11:05,2.2
38.2285213303,59.9341643994,2017-09-04 00:00:05,3.0,$^b$2017-09-27 22:00:05,3.4,2017-11-25 18:11:05,3.4
39.0871950358,49.8480561921,2017-08-30 23:26:05,1.7,2017-09-23 21:26:05,2.0,2017-11-18 18:37:05,2.4
39.1191462643,62.6704841545,2017-09-04 00:26:05,2.6,2017-09-27 22:26:05,2.7,2017-11-25 18:37:05,3.1
40.0580565398,52.5607257898,2017-08-30 23:26:05,1.6,2017-09-23 21:26:05,1.7,2017-11-18 18:37:05,2.2
40.1700706101,65.4068039096,2017-09-04 00:26:05,2.1,2017-09-27 22:26:05,2.6,2017-11-25 18:37:05,2.9
41.2668652502,55.4118387671,2017-09-04 00:00:05,2.3,2017-09-27 22:00:05,2.3,2017-11-25 18:11:05,2.7
41.738945248,68.0958233498,2017-09-03 01:00:05,1.7,2017-09-26 21:11:05,1.7,2017-11-24 18:11:05,2.4
42.1574901842,58.1481585222,2017-09-04 00:00:05,2.7,2017-09-27 22:00:05,3.5,2017-11-25 18:11:05,2.8
43.3509962819,70.8084929475,2017-09-03 01:00:05,1.9,2017-09-26 21:11:05,1.8,2017-11-24 18:11:05,2.8
43.7552718262,60.8844782774,2017-09-04 00:26:05,2.3,2017-09-27 22:26:05,2.5,2017-11-25 18:37:05,2.8
44.26788367,76.7963348687,2017-09-03 01:26:05,1.9,2017-09-26 21:37:05,1.6,2017-11-24 18:37:05,3.1
44.8061961721,63.6207980325,2017-09-04 00:26:05,2.0,2017-09-27 22:26:05,2.3,2017-11-25 18:37:05,2.5
44.9375,79.57222222,2017-09-03 01:26:05,2.2,2017-09-26 21:37:05,1.8,2017-11-24 18:37:05,3.8
47.2302167921,66.218674408,2017-09-03 01:00:05,2.1,2017-09-26 21:11:05,1.6,2017-11-24 18:11:05,2.5
48.8422678259,68.9313440057,2017-09-03 01:00:05,1.9,2017-09-26 21:11:05,1.6,2017-11-24 18:11:05,2.8
57.884694329,78.0793181194,2017-09-03 01:26:05,2.1,2017-09-26 21:37:05,1.8,2017-11-24 18:37:05,3.6
58.554310659,80.8552054707,2017-09-03 01:26:05,1.6,2017-09-26 21:37:05,1.3,2017-11-24 18:37:05,2.2
\end{filecontents*}
\csvreader[tabular=llrc|rc|rc,
    head=false,
    table head=\toprule,
    late after line=\\,
    late after first line=\\\midrule,
    table foot=\bottomrule,
    ]{obs_table_g299232.csv}{}{\csvcoli & \csvcolii & \csvcoliii & \csvcoliv & \csvcolv &\csvcolvi & \csvcolvii & \csvcolviii}
\clearpage
    \centering
    \captionof{table}{Differential source counts corresponding to Figure \ref{fig:rms}. From left to right: bin range of flux densities, bin centre flux density, number of sources in bin, correction factor to account for varying sensitivity across the field, and the final Euclidean normalized source density. The uncertainties were calculated following Poisson statistics.
    \label{tab:sources}}
    \begin{filecontents*}{normalized_source_counts.csv}
$dF_{\nu}$\,(Jy),$F_{\nu}$\,(Jy),Raw counts,Completeness weight,Normalized counts (Jy$^{1.5}$\,sr$^{-1}$)
0.0051--0.0056,0.0054,$84\pm9$,141.00,$29\pm3$
0.0056--0.0062,0.0059,$117\pm10$,80.57,$27\pm2$
0.0062--0.0068,0.0065,$241\pm15$,37.60,$30\pm1$
0.0068--0.0075,0.0072,$380\pm19$,23.50,$34\pm1$
0.0075--0.0083,0.0079,$582\pm24$,16.59,$42\pm1$
0.0083--0.0091,0.0087,$850\pm29$,10.25,$44\pm1$
0.0091--0.010,0.0095,$1177\pm34$,6.96,$48\pm1$
0.010--0.013,0.012,$5831\pm76$,4.15,$64.5\pm0.8$
0.013--0.018,0.016,$9784\pm98$,2.29,$92.1\pm0.9$
0.018--0.024,0.021,$12850\pm113$,1.63,$132\pm1$
0.024--0.032,0.028,$14745\pm121$,1.22,$175\pm1$
0.032--0.042,0.039,$16235\pm127$,1.05,$256\pm2$
0.042--0.056,0.049,$15433\pm124$,1.01,$359\pm2$
0.056--0.075,0.066,$14223\pm119$,1.00,$505\pm4$
0.075--0.10,0.087,$11295\pm106$,1.00,$618\pm5$
0.10--0.13,0.12,$9692\pm98$,1.00,$817\pm8$
0.13--0.18,0.16,$8629\pm92$,1.00,$1120\pm12$
0.18--0.24,0.21,$6627\pm81$,1.00,$1325\pm16$
0.24--0.32,0.28,$5938\pm77$,1.00,$1828\pm23$
0.32--0.42,0.37,$4352\pm65$,1.00,$2064\pm31$
0.42--0.56,0.49,$3339\pm57$,1.00,$2438\pm42$
0.56--0.75,0.66,$2391\pm48$,1.00,$2689\pm54$
0.75--1.0,0.87,$1914\pm43$,1.00,$3315\pm75$
1.0--1.3,1.2,$1521\pm39$,1.00,$4056\pm104$
1.3--1.8,1.6,$773\pm27$,1.00,$3175\pm114$
1.8--2.4,2.1,$619\pm24$,1.00,$3915\pm157$
2.4--3.2,2.8,$320\pm17$,1.00,$3116\pm174$
3.2--4.2,3.7,$200\pm14$,1.00,$2999\pm212$
4.2--5.6,4.9,$189\pm13$,1.00,$4365\pm317$
5.6--7.5,6.6,$55\pm7$,1.00,$1956\pm263$
7.5--10.0,8.8,$61\pm7$,1.00,$3341\pm427$
10.0--16.2,13.1,$37\pm6$,1.00,$2243\pm368$
16.2--26.2,21.2,$12\pm3$,1.00,$1499\pm432$
26.2--42.5,34.3,$18\pm4$,1.00,$4633\pm1092$
42.5--68.7,55.6,$2\pm1$,1.00,$1060\pm750$
\end{filecontents*}
\csvreader[tabular=lllcl,
        head=false,
        table head=\toprule,
        late after line=\\,
        late after first line=\\\midrule,
        table foot=\bottomrule,
        ]{normalized_source_counts.csv}{}{\csvcoli & \csvcolii & \csvcoliii & \csvcoliv & \csvcolv}

\raggedright
\bsp	
\label{lastpage}
\end{document}